\documentclass{article}

\usepackage{arxiv}

\usepackage[utf8]{inputenc} 
\usepackage[T1]{fontenc}    
\usepackage{hyperref}       
\usepackage{url}            
\usepackage{booktabs}       
\usepackage{amsfonts}       
\usepackage{nicefrac}       
\usepackage{microtype}      
\usepackage{lipsum}		
\usepackage{graphicx}
\usepackage{natbib}
\usepackage{doi}

\usepackage{amsmath}

\usepackage{amsmath}
\usepackage{amssymb}  
\usepackage{amsthm}  
\usepackage{mathrsfs}
\usepackage{algorithm}
\usepackage{algorithmic}
\usepackage{dsfont}
\usepackage{xcolor}
\usepackage{multirow}

\usepackage{amsthm}  
\usepackage{mathrsfs}

\usepackage{enumerate}
\usepackage{enumitem}

\usepackage{array}
\newcolumntype{L}[1]{>{\raggedright\let\newline\\\arraybackslash\hspace{0pt}}m{#1}}
\newcolumntype{R}[1]{>{\raggedleft\let\newline\\\arraybackslash\hspace{0pt}}m{#1}}
\newcolumntype{C}[1]{>{\centering\let\newline\\\arraybackslash\hspace{0pt}}m{#1}}

\newcommand{\myparbox}[2]{\parbox{#1}{\vspace{.3\baselineskip}\raggedright #2 \vspace{.3\baselineskip}}}
\newcommand{\myCparbox}[2]{\parbox{#1}{\vspace{.3\baselineskip}\centering #2 \vspace{.3\baselineskip}}}

\usepackage[caption=false,font=footnotesize]{subfig}

\newcommand{\LCOMMENT}[1]{/*~#1~*/}

\newcommand{\syn}{\text{Syn}}

\newcommand{\h}{\text{\#}h}

\newcommand{\bv}{{\bf v}}

\newcommand{\thes}{\mathscr{T}}

\newcommand{\hr}{\text{hit-ratio}}
\newcommand{\hrRE}{\text{\#REval-hit-ratio}}

\newcommand{\RG}{\langle \mathscr{R}, \mathscr{G} \rangle}

\newcommand{\hockey}{\text{\#hockey}}
\newcommand{\bowling}{\text{\#bowling}}
\newcommand{\golf}{\text{\#golf}}
\newcommand{\sport}{\text{\#sport}}
\newcommand{\championship}{\text{\#championship}}
\newcommand{\champion}{\text{\#champion}}
\newcommand{\winner}{\text{\#winner}}
\newcommand{\tournament}{\text{\#tournament}}
\newcommand{\football}{\text{\#football}}
\newcommand{\soccer}{\text{\#soccer}}
\newcommand{\footy}{\text{\#footy}}
\newcommand{\rugby}{\text{\#rugby}}
\newcommand{\sports }{\text{\#sports}}
\newcommand{\exercise}{\text{\#exercise}}
\newcommand{\keepfit}{\text{\#keeepfit}}
\newcommand{\swim}{\text{\#swim}}
\newcommand{\walking}{\text{\#walking}}
\newcommand{\yoga}{\text{\#yoga}}

\newcommand{\dive}{\text{\#dive}}
\newcommand{\paddle}{\text{\#paddle}}

\usepackage{tikz, pgf, pgfplots}
\usetikzlibrary{shapes.geometric}
\usetikzlibrary{arrows.meta}
\usetikzlibrary{calc}
\usetikzlibrary{plotmarks}

\definecolor{airforceblue}{rgb}{0.36, 0.54, 0.66}

\title{\#REval: A Semantic Evaluation Framework for Hashtag Recommendation}

\author{
 Areej Alsini \\
  Umm Al-Qura University\\
  \texttt{aosini@uqu.edu.sa} \\
   \And
 Du Q. Huynh \\
 The University of Western Australia\\
  \texttt{du.huynh@uwa.edu.au} \\
  \And
 Amitava Datta \\
 The University of Western Australia\\
  \texttt{amitava.datta@uwa.edu.au} \\
}

\renewcommand{\shorttitle}{\textit{arXiv} Template}

\hypersetup{
pdftitle={A template for the arxiv style},
pdfsubject={q-bio.NC, q-bio.QM},
pdfauthor={David S.~Hippocampus, Elias D.~Striatum},
pdfkeywords={First keyword, Second keyword, More},
}

\begin{document}
\maketitle

\begin{abstract}
Automatic evaluation of hashtag recommendation models is a fundamental task in many online social network systems. In the traditional evaluation method, the recommended hashtags from an algorithm are firstly compared with the ground truth hashtags for exact correspondences. The number of exact matches is then used to calculate the hit rate, hit ratio, precision, recall, or F1-score. This way of evaluating hashtag similarities is inadequate as it ignores the semantic correlation between the recommended and ground truth hashtags. To tackle this problem, we propose a novel semantic evaluation framework for hashtag recommendation, called \#REval. This framework includes an internal module referred to as \textit{BERTag}, which automatically learns the hashtag embeddings. We investigate on how the \#REval framework performs under different word embedding methods and different numbers of
synonyms and hashtags in the recommendation using our proposed \hrRE\ measure. Our experiments of the proposed framework on three large datasets show that \#REval gave more meaningful hashtag synonyms for hashtag recommendation evaluation. Our analysis also highlights the sensitivity of the framework to the word embedding technique, with \#REval based on BERTag more superior over \#REval based on FastText and Word2Vec.
\end{abstract}


\section{Introduction}
\label{sec:introduction}
It is important to have frameworks that can automatically and accurately evaluate the performances of algorithms. In the hashtag recommendation research area, algorithms that can help recommend hashtags to the user while the tweet is being written have received a great amount of attention in recent years. This is due to the increased numbers of online messaging systems and online users every year. Hashtags are innovative social identifiers prefixed with the \# sign. They work as annotations to their corresponding tweets. They enable users to quickly find tweets of a topic in mind and they help to promote engagement among users. Although hashtag recommendation models have received considerable interest due to the increased popularity of social media platforms~\cite{Alsini-2021}, studies on their evaluation methods are still in their infancy. 

Current evaluation methods for hashtag recommendation mainly focus on the \textit{accuracy-based metrics}~\cite{Kywe-2012,Zhao-2016,Li-2016,Alsini-2017,Kowald-2017,Zhang-2017,Shi-2018,Alsini-2020,alsini2020hit} such as \textit{hit rate}, \textit{hit ratio}~\cite{alsini2020hit}, \textit{precision}, \textit{recall}, and \textit{F1} scores. These metrics are dependent on the exact matching of the recommended hashtags with the actual list of hashtags used in the tweets of the user. For evaluation purpose, we refer to this actual list of hashtags as the \textit{ground truth} hashtags. Some other researchers used human evaluation~\cite{Mazzia-2011} to assess their hashtag recommendation models because the suggested hashtags did not match the ground truth hashtags even though they were relevant.

Evaluating hashtag recommendations is a very challenging task, as illustrated in the following tweets taken from the \#Covid19 thread in Twitter:
\begin{enumerate}[label={(\arabic*)},nolistsep]
\item \textit{Know the facts about \#Covid19...Don’t panic \#StayHomeSA \#StaySafe \#21DayLockdown}.
\item \textit{\#Covid19: A Crucial Impact On \#Defence \#Industry}.
\item \textit{\#Covid19 Together Let's \#FightCovid19 \#FlattenTheCurve \#StopTheSpread}.
\item \textit{Please list \#covid19 helpline numbers various states in the U.S. \#covid2019 \#corona \#coron…}
\end{enumerate}
These tweets use 2-4 hashtags. For example in tweet (1), the ground truth hashtags are \textit{\#Covid19}, \textit{\#StayHomeSA}, \textit{\#StaySafe}, and \textit{\#21DayLockdown}. It is clear from these examples that the hashtags used in all these tweets share similar meaning with the hashtags \textit{\#Covid\_19} and \textit{\#coronavirus}. However, if a recommendation model recommended these two hashtags, then the traditional method of evaluation, which looks for exact matches of the character strings, would deem them as incorrect.

Another challenge is that these hashtags are \textit{user-generated} contents which make them prone to spelling mistakes. Some hashtags might be connected mistakenly with other words or punctuation if no space is left after writing the hashtag. This is shown clearly in the hashtags \textit{\#Covid19...Don’t}, \textit{\#Covid19:} and \textit{\#coron…} used in tweets (1), (2) and (4), respectively. With these hashtags in the ground truth and the use of the traditional method of evaluation, hashtag recommendation models cannot be evaluated properly.

Thus, there is a need for a semantic evaluation method that can evaluate hashtag recommendation models based on the semantic correlation of the recommended hashtags with their corresponding ground truth ones. Our research contributions for tackling the challenges described above are summarised below:

\begin{itemize}
\item We propose \#REval, a semantic evaluation framework that evaluates the performance of hashtag recommendation models by incorporating hashtag synonyms. Based on the hit ratio used in~\cite{alsini2020hit}, we introduce a new measure referred to as \textit{\hrRE} for synonym evaluation.
\item We propose a method that automatically constructs a \textit{thesaurus} of hashtags where every hashtag is associated with a set of hashtag synonyms that share similar meanings. 
\item We propose a hashtag embedding module called \textit{BERTag}, which encodes hashtags using the tweet embeddings generated by the transformer-based model \textit{BERTweet}~\cite{nguyen-etal-2020-bertweet}. Our encoded hashtags are represented together with tweets in the same semantic space. 
\item To evaluate the performance of our BERTag embeddings, we compare them with two popular word embedding techniques, namely \textit{Word2Vec} and \textit{FastText}, on three large datasets, two of which are general and the remaining one is domain-specific. Our evaluation results show that the BERTag embeddings give more relevant hashtag synonyms.

\end{itemize}

The paper is organised as follows. Section~\ref{review} reviews studies related to semantic evaluation methods in text-based problems, domain modelling and synonym identification, hashtag representations, and transformer-based models. Our proposed framework \#REval is detailed in Section~\ref{Our_Proposed_Method}. Experimental setups, details of the datasets, results, and discussion are presented in Section~\ref{sec:Experiments}. 
Finally, concluding remarks and future work are given in Section~\ref{Conclusion}.

\section{Literature Review}
\label{review}

\subsection{Semantic Evaluation Methods for Text Analysis}

There are various automatic language-based evaluation metrics. In the context of machine translation, BiLingual Evaluation Understudy (BLEU)~\cite{Papineni-2002} is an automatic evaluation method. It matches the n-gram vectors of a given translated sentence (known as \textit{candidate}) with the n-gram vectors of multiple human translations (known as \textit{references}) to yield a weighted geometric mean of all the n-gram precisions. Thus, the more is the number of matches, the higher is the BLEU score. METEOR~\cite{banerjee-2005} is another automatic method for evaluating the outputs of machine translation algorithms. It gives further improvement to BLEU by using additional elements, such as stemming and synonym matching, in the evaluation process. While BLEU scores are defined based on precisions only, METEOR scores are defined as the harmonic means of unigram precisions and recalls, with a heavier weight on the latter. In the context of text summarization, Recall-Oriented Understudy for Gisting Evaluation (ROUGE)~\cite{lin-2004} is introduced as a set of automatic metrics, which evaluate the quality of a candidate summary by matching it to the references created by humans using the n-grams, word sequences, and word pairs. 

Evaluation metrics have also gained increasing interest in the image captioning research area. CIDEr~\cite{VedantamZP-2015} and SPICE~\cite{Anderson-2016} are image captioning evaluation methods that incorporate human consensus in the description of an image. Precision and recall are considered when estimating the similarity between the candidate sentence and the sentences annotated by humans. 
CIDEr measures the similarity between the n-gram of words occurring in the candidate sentence and all the reference sentences. SPICE~\cite{Anderson-2016}, on the other hand, constructs a semantic parsing that converts the captions into scene graphs. It has been shown to outperform all previously mentioned metrics. An interesting finding is that the more is the number of reference sentences, the higher is the SPICE score. 

In all the evaluation methods mentioned above, human references were considered as the base of evaluation. For hashtag recommendation, the ground truth hashtags, which are the hashtags present in the test tweets, are used as references for evaluation. In this regard, the only available method is to perform exact matching between the recommended hashtags and the hashtags from the ground truth. In the literature, accuracy-based metrics such as hit rate, precision, recall, hit ratio, and F1-score are commonly used for evaluating hashtag recommendation models~\cite{alsini2020hit}: Hit rate considers the recommendation as a hit if at least a single recommended hashtag has an identical match with a hashtag from the ground truth; precision and recall measure the proportion of correctly matched hashtags in the recommended hashtags and in the ground truth hashtags, respectively; hit ratio calculates the ratio of the hits against the minimum number of hashtags in the recommendation and in the ground truth; F1-score is a harmonic mean of precision and recall. A shortcoming of these metrics is that they do not take into consideration the semantic correlation in the evaluation of hashtag recommendation. The aim of our proposed framework is to overcome this shortcoming.

\subsection{Domain Modeling and Synonym Identification}
Domain modelling schemes represent data in one of the following structured form~\cite{Gilchrist-2003}: \textit{Taxonomy}, \textit{ontology}, and \textit{thesaurus}. In \textit{taxonomy}, multiple levels of concept generality are represented using a hierarchical tree structure, with generic terms at the top of the tree to specific terms at the bottom; in \textit{ontology} of a given domain, of interest is to retrieve a formal and explicit specification of the shared conceptualisation; in \textit{thesaurus}~\cite{Schwarzdissertation-2005}, every term has a reference to all the other terms that have similar meanings to it. The relationship of a concept term with its synonyms in a thesaurus can be \textit{hierarchy}, \textit{association}, or \textit{equivalence}~\cite{Schwarzdissertation-2005}. Synonym expansion has been used in many applications, such as information retrieval~\cite{Al-Khateeb-2017} and search-related domains~\cite{boteanu-2019}. Botenanu et al.~\cite{boteanu-2019} enhanced the search for Amazon products via building large taxonomies of shopping items. 

In the literature, three main methods have been used to identify synonyms: WordNet~\cite{boteanu-2019}, Named Entity Recognition~\cite{Bohn-2010,Sotomayor-2017}, and word embedding~\cite{Wang-2015-medical, Leeuwenberg-2016, Landthaler-2017,Ali-2019,Cheng-2019-multi}. Handler et al.~\cite{Handler-2014} highlighted that word embedding generates a larger number of synonyms compared to other methods. Hazem et al.~\cite{hazem-2018} used word embedding to identify synonyms for multi-word terms and Ali et al.~\cite{Ali-2019} used word embeddings to distinguish between synonyms and antonyms of the embeddings. Both papers used Word2Vec as their word embedding technique.

In our paper, our interest is to associate hashtags that have similar
meanings. So we investigate various word embedding techniques to
represent our hashtags and to build our thesaurus.

\subsection{Hashtag Representations}
Hashtags have been used to promote TV-shows~\cite{Michele-2019} and utilised by health care professionals to share ideas and policies~\cite{Ojo-2021}. Being able to semantically compare
hashtags help to quickly identify social behaviours in the digital world, such as bullying~\cite{Calvin-2015} and activism~\cite{Virginia-2021}, and aggregate disaster responses~\cite{Jishnu-2020}. Generally, machine learning tasks require the hashtags to be represented as feature vectors (embeddings or representations). However, learning the embeddings of hashtags can be very challenging. As hashtags are user-generated, problems such as sparsity, polysemy and synonymy are common~\cite{LiuHe-2018}. Hashtags can be written as acronyms and can include numbers, misspelled words, connected words, and connected words separated by an underscore and/or full stops. Besides, a hashtag can have multiple meanings, and multiple hashtags can share the same meaning.

Hashtags have been dealt with differently in the literature. Some researchers assume that hashtags are independent components and should not be treated as words~\cite{Weston-2014,LiLiuHu-2019,ZhoYang-2020}. Thus, they preserve hashtags as independent tokens without pre-processing them. Some other researchers assume that the same peculiarities are shared between words in tweets and hashtags. Accordingly, they treat words and hashtags similarly after removing the punctuation, including the \# sign. Hashtags are then tokenised into words and sub-words using a tokenisation method.

Various methods have been proposed to learn hashtag representations, such as Bag of Words methods (BOW)~\cite{Tsur-2012,Tsur-2013,Javed-2018}, topic models~\cite{Zhao-2016}, graph models~\cite{LiuHe-2018}, and word embedding techniques~\cite{Dey-2017}. Tsur et al.~\cite{Tsur-2012,Tsur-2013} and Javed et al.~\cite{Javed-2018} aggregated tweets containing the same hashtag as a document. They then represented each hashtag using BOW of the most frequent words in each document. Zhao et al.~\cite{Zhao-2016} proposed Hashtag-LDA to generate joint representations of hashtags and words in tweets to discover latent topics and global hashtags. Dey et al.~\cite{Dey-2017} proposed EmTaggeR that treats hashtags as words in the tweets. They pre-trained a skip-gram Word2Vec model over tweets to derive the embeddings of hashtags. Liu et al.~\cite{LiuHe-2018} proposed Hashtag2Vec that creates a heterogeneous hierarchical graph with hashtags, tweets, and words of tweets to derive the embeddings of hashtags.  

As hashtags are \textit{user generated} and the free style of creating hashtags change over time, applying the previous methods can be difficult. BOW methods focus on the presence and absence of words in a set of tweets containing hashtags and ignore the contextual meaning. LDA-based models adopt statistical methods where latent parameters need to be recalculated with new data. 
Using the graph models, the network graph structure can change significantly with the dynamic change of topics over time. 
Word embedding techniques such as Word2Vec and FastText are trained on the word-level, and pre-trained Word2Vec and FastText models can perform poorly on Twitter data~\cite{Stewart-2019}. Most researchers re-trained their word embedding models, especially when working on domain-specific Twitter data~\cite{Alsini-2019,Alsini-2020}.

\subsection{Transformer-based Models}

The context vector of a word can be encoded with respect to its surrounding words using an attention mechanism such as the transformer~\cite{Vaswani-2017}. In 2018, Google's \textit{Bidirectional Encoder Representations from Transformers} (BERT)~\cite{devlin-etal-2019-bert} was a breakthrough in the NLP domain. BERT is trained on extensive corpora using the masked words for predicting the next sentence. BERT comes with two model sizes:
\begin{itemize}
    \item BERT-base, which contains 12 layers (transformer blocks), 12 attention heads, and 110 million parameters;
    \item BERT-Large, which includes 24 layers, 16 attention heads and 340 million parameters.
\end{itemize}

Recently, various transformer-based models, such as RoBERTa~\cite{liu2019-roberta}, AlBERT~\cite{lan2020albert}, and BERTweet~\cite{nguyen-etal-2020-bertweet},  have been proposed. 
RoBERTa optimizes BERT by training it on larger corpora and more training iterations with different masking patterns. AlBERT optimizes BERT by using fewer parameters and thus shortens the training time. BERTweet, on the other hand, uses a similar architecture as BERT-base but was trained using RoBERTa's pre-training procedure on Twitter data. BERTweet outperformed RoBERTa and XLM-R on two classification datasets.

BERTweet-base was pre-trained on 850 million English tweets collected from 01/2012 to 08/2019. Pre-trained transformer-based models enable researchers to achieve better results with minimal fine-tuning on various NLP tasks such as classification~\cite{antoun-2020}, Named Entity Recognition~\cite{Chen-2020}, and Question-Answering~\cite{Kayesh-2020}. 
All the previously mentioned transformer-based models generate embeddings at the sentence level. Unlike all the above transformer-based models which generate word embeddings at the sentence level, our module, BERTag, encodes hashtags using a pre-trained BERTweet model.

\section{The \#REval Framework}
\label{Our_Proposed_Method}
Given a dataset $\mathscr{D}\!=\!\{(t_{1},H_{1}),(t_{2},H_{2}),\cdots\}$ that contains a list of tweets and their corresponding hashtags where $H_{i}\!=\!\{h_1,h_2,\cdots\}$, let  $\{(\mathscr{R},\mathscr{G})\}\!=\!\{(R_1,G_1),(R_2,G_2),\cdots \}$ be the collection of tuples representing the test set obtained from a hashtag recommendation model. Here, $R_i=\{\hat{h}'_1,\hat{h}'_2,...\}$ is the list of the top-$r$ recommended hashtags (for a user-defined $r$ value) of the $i^{\text{th}}$ test tweet and $G_i=\{h'_1,h'_2,...\}$ is the corresponding ground truth hashtags. Our goal is to measure the performance of a given hashtag recommendation model by computing the average \#REval-hit-ratio (described later in Section~\ref{sec:hrRE}), taking into account the semantic correlation between hashtags. The notations used in the rest of the paper are listed in Table~\ref{tab:notations}. 

Our \#REval framework is designed to detect the semantic correspondences of the lists of hashtags between $\mathscr{R}$ and $\mathscr{G}$. It comprises the following modules: \textit{BERTag, synonym and thesaurus construction}, and \textit{semantic evaluation}. These modules are detailed in the subsections below.

\begin{table}
\centering
{\renewcommand{\arraystretch}{1.4}
\setlength{\tabcolsep}{1pt}
\caption{Notations used in the paper.}
\label{tab:notations}
\scalebox{0.8}{
\begin{tabular}{| L{0.35\textwidth} | L{.87\textwidth} |}
\hline
$\mathscr{D}\!=\!\{(t_1,H_1),(t_2,H_2),... \}$  & the processed dataset that contains a list of hashtagged tweets $t_{i}$, each of which has a corresponding list of hashtags $H_{i}$. \\ \hline
$T\!=\!\{t_1,t_2,...\}$ & the set of all tweets. \\ \hline
$h$; $h_i$  & an arbitrary hashtag; the $i^{\text{th}}$ hashtag in a set. \\ \hline
${(\eta_{t},h)}$ & the index $\eta_{t}$ of a tweet $t$ for the hashtag $h$. \\ \hline
$m$ & total number of clusters (i.e., total number of unique hashtags in the dataset). \\ \hline
$\bv_t$; $\bv_h$ & embedding for tweet $t$; embedding for hashtag $h$ \\ \hline
$n_h$ & total number of tweets having $h$ as a hashtag. \\ \hline
$R_i\!=\!\{h_1, h_2,...,h_r\}$ & the list of top-$r$ (for some positive integer $r$) recommended hashtags for the $i^{\text{th}}$ test tweet. \\ \hline 
$G_i\!=\!\{h'_1, h'_2,...,h'_{n_i}\}$ & the ground truth list of hashtags for the $i^{\text{th}}$ test tweet; the length $n_i$ of the list varies from tweet to tweet. \\ \hline
$\RG \!=\! \{(R_1,G_1), ...,(R_n,G_n)\}$ & the test set for evaluation, where $R_i$ is to be verified against $G_i$. \\ \hline
$\syn_k(h)$ & the set containing hashtag $h$ and the $k$ closest synonyms of $h$, i.e., the set has $k+1$ elements. \\ \hline
$\mathscr{E}\!=\!\{(h,\bv_{h}) \,|\, \forall h\}$ & a dictionary of hashtags and their embeddings. \\ \hline
$\thes \!=\! \{(h,\syn(h)) \,|\, \forall h\}$  & a thesaurus of hashtags and their synonyms. \\ \hline
\end{tabular}}}
\end{table}

\subsection{BERTag}
\label{sec:bertag}
Let $T=\{t_i\, | \, \forall i \}$ be the set containing all the tweets in $\mathscr{D}$. We consider the hashtags as labels for tweets, where every tweet has a single hashtag label. For a tweet with multiple hashtags, we duplicate the tweet a sufficient number of times to match the number of hashtags. For example, if a tweet $t_j \in T$ contains two hashtags $\{h_a, h_b\}$, for some integers $a$ and $b$, we create the pairs $(t_j^{(1)},h_a)$ and $(t_j^{(2)},h_b)$. The superscript of each $t_j$ denotes the duplication of the tweet to match the number of hashtags. The absolute ordering of tweets for a given hashtag is not important. The subscripts under $h$ correspond to the unique ID of each distinct hashtag. 
Retweets with the same label are removed. For each hashtag $h$, we also keep the index $\eta_{t}$ of the corresponding tweet $t$, i.e., we store $\{(\eta_{t},h)\,|\, \forall t\}$, for future retrieval of all the tweets for any given hashtag $h$. 

Tweets with their hashtags are the input to our module BERTag. The output of BERTag is a dictionary $\mathscr{E}=\{(h,\bv_{h})\,|\, \forall h \}$, where each dictionary item is a hashtag $h$ and its semantic representation $\bv_h$. 
To generate $\mathscr E$, three stages are involved in BERTag: \textit{pre-processing}, \textit{extracting tweet embeddings}, and \textit{computing hashtag embeddings}. 

\begin{figure}[th!]
\begin{center}
\resizebox{0.8\columnwidth}{!}{
\begin{tikzpicture}
[box/.style={rectangle, draw=black, fill=none, align=center},
roundedBox/.style={rectangle, rounded corners, draw=black, align=center},
blank/.style={rectangle, draw=none},
smallarrow/.style={-{Stealth[length=2mm, width=2mm]}},
arrow/.style={-{Stealth[length=3mm, width=2mm]}}]

\makeatletter
\newcommand{\gettikzxy}[3]{%
  \tikz@scan@one@point\pgfutil@firstofone#1\relax
  \edef#2{\the\pgf@x}%
  \edef#3{\the\pgf@y}%
}
\makeatother

\node[box, minimum width=6.8cm, minimum height=0.55cm, line width=0.2mm, rotate=90] (preprocessing) 
  at (1.5,0.4) {\textbf{Pre-processing}};
\node[roundedBox, fill=airforceblue!80, minimum width=4.5cm] (pretrain)
  at (5.2,2.5) {\textbf{Pre-trained tokenizer}};
\node[roundedBox, rounded corners=20, fill=airforceblue!80, minimum width=4.5cm, minimum height=4.2cm] 
  (bluebox) at (5.2,-0.3) {};
\node[roundedBox, fill=blue!5, rotate=90] (transf1) at (5.1,-0.3) {~~Transformer block~~};
\node[roundedBox, fill=blue!5, rotate=90] (transf2) at (5.9,-0.3) {~~Transformer block~~};
\node[roundedBox, fill=blue!5, rotate=90] (transf3) at (6.9,-0.3) {~~Transformer block~~};

\node [rounded corners=10, fill=none, draw=black, minimum width=5.0cm, minimum height=5.9cm,
    line width=0.2mm] (BERTweet)  at (5.2, -0.05) {};
\node[anchor=south, align=center] at (BERTweet.south)
  {\textbf{Pre-trained BERTweet}};

\node[box,minimum width=4cm, minimum height=6.8cm, line width=0.2mm] (semantic)
  at (10.5,0.4) {};
\node[anchor=south, text width=6cm, align=center] at (semantic.south)
  {\scriptsize Semantic space \\[-1ex] (hashtag embedding space)};

\draw [smallarrow] ( $(transf1.north) + (-0.5,1.3)$ ) -- ( $(transf1.north) + (0.0,1.3)$ )
  node[above, left, blank] { {\scriptsize [CLS]}~~~~~};
\draw [smallarrow] ( $(transf1.north) + (-0.5,0.8)$ ) -- ( $(transf1.north) + (0.0,0.8)$ )
  node[above, left, blank] { {\scriptsize reading}~~~~~};
\draw [smallarrow] ( $(transf1.north) + (-0.5,0.3)$ ) -- ( $(transf1.north) + (0.0,0.3)$ )
  node[above, left, blank] { {\scriptsize nice}~~~~~};
\draw [smallarrow] ( $(transf1.north) + (-0.5,-0.2)$ ) -- ( $(transf1.north) + (0.0,-0.2)$ )
  node[above, left, blank] { {\scriptsize books}~~~~~};
\draw [smallarrow] ( $(transf1.north) + (-0.5,-0.7)$ ) -- ( $(transf1.north) + (0.0,-0.7)$ )
  node[above, left, blank] { {\scriptsize today}~~~~~};
\draw [smallarrow] ( $(transf1.north) + (-0.5,-1.2)$ ) -- ( $(transf1.north) + (0.0,-1.2)$ )
  node[above, left, blank] { {\scriptsize [SEP]}~~~~~};

\draw [-] (transf1.south) -- (transf2.north);
\node[blank] at ( $(transf2.south) + (0.4,0)$ ) {...~~~~};
\draw [arrow] (pretrain.south) -- (bluebox.north);

\foreach \value/ \i / \h in {2.2/1/a, 1.5/2/b, 0.8/3/c}
    \draw [smallarrow] ( $(preprocessing.north) + (-0.4,\value)$ ) -- 
        ( $(preprocessing.north) + (0,\value)$ )
         node[left,midway] {\footnotesize $(t_1^{(\i)},h_{\h})$~~};
         
\draw [smallarrow] ( $(preprocessing.north) + (-0.4,-0.4)$ ) -- 
        ( $(preprocessing.north) + (0,-0.4)$ )
         node[left,midway] {\footnotesize $\vdots$~~~~~~};    
         
\draw [smallarrow] ( $(preprocessing.north) + (-0.4,-1.2)$ ) -- 
        ( $(preprocessing.north) + (0,-1.2)$ )
         node[left,midway] {\footnotesize $(t_i^{(j)},h_{d})$~~};    
\draw [smallarrow] ( $(preprocessing.north) + (-0.4,-2.0)$ ) -- 
        ( $(preprocessing.north) + (0,-2.0)$ )
         node[left,midway] {\footnotesize $\vdots$~~~~~~};    

\draw [arrow] ( $(preprocessing.south) + (0,2.9)$ ) -- 
        ( $(semantic.west) + (0,2.9)$ )
         node[above,midway] {\footnotesize $\{(\eta_{t},h) \,|\, \forall h\}$~~~~~~}; 

\gettikzxy{(preprocessing.south)}{\px}{\py}
\gettikzxy{(BERTweet.west)}{\qx}{\qy}

\foreach \value/ \i in {1.2/1, 0.5/2, -0.2/3}
    \draw [smallarrow] ( $(\px, \qy) + (0,\value)$ ) -- ( $(\qx, \qy) + (0, \value)$ )
         node[above,midway] {\footnotesize $t_{\i}$};

\draw [smallarrow] ( $(\px, \qy) + (0,-1.8)$ ) -- ( $(\qx, \qy) + (0, -1.8)$ )
     node[above,midway] {\footnotesize $t_i$};
\draw [draw=none] ( $(\px,\qy) + (0,-1.8)$ )  rectangle ( $(\qx,\qy) + (0,0.4)$ )  
    node[midway] {$\vdots$};
\draw [draw=none] ( $(\px,\qy) + (0,-2.6)$ )  rectangle ( $(\qx,\qy) + (0,-1.8)$ )  
    node[midway] {$\vdots$};

\gettikzxy{(BERTweet.east)}{\qx}{\qy}
\gettikzxy{(semantic.west)}{\rx}{\ry}

\foreach \value/ \i in {1.2/1, 0.5/2, -0.2/3}
    \draw [smallarrow] ( $(\qx, \qy) + (0,\value)$ ) -- ( $(\rx, \qy) + (0, \value)$ )
         node[above,midway] {\footnotesize $\bv_{t_{\i}}$};

\draw [smallarrow] ( $(\qx, \qy) + (0,-1.8)$ ) -- ( $(\rx, \qy) + (0, -1.8)$ )
     node[above,midway] {\footnotesize $\bv_{t_{i}}$};
\draw [draw=none] ( $(\qx,\qy) + (0,-1.8)$ )  rectangle ( $(\rx,\qy) + (0,0.4)$ )  
    node[midway] {$\vdots$};
\draw [draw=none] ( $(\qx,\qy) + (0,-2.6)$ )  rectangle ( $(\rx,\qy) + (0,-1.8)$ )  
    node[midway] {$\vdots$};
    
\draw [arrow] (semantic.east) -- ( $(semantic.east) + (0.6,0)$ )   
    node[right] {$\mathscr{E}$};

\pgfmathsetseed{123}


\coordinate (p1) at (11.5, 2.8);
\coordinate (p2) at (11, 0.7);
\coordinate (p3) at (9.5, 1.2);
\coordinate (p4) at (10, 3);
\coordinate (p5) at (9.6, -1.0);

\foreach \point/\colour/\N/\sigma/\index in {p1/orange/20/0.75/1, p2/blue!50/18/0.7/4, 
                        p3/teal/15/0.6/3,  p4/green/30/0.7/2,
                        p5/red/20/0.75/m} {
    \foreach \i in {1,...,\N} {
        \coordinate (p) at ( $(\point) + (rand*\sigma, rand*\sigma)$ );
        \filldraw[\colour] (p) circle (2pt);
    }; 
    \draw[mark=x, mark size=3pt, mark options={color=black}, line width=1.5] plot coordinates {(\point)}
        node [right] {$h_{\index}$};
}

\end{tikzpicture}
}
\end{center}
\caption{The BERTag Architecture. The inputs are the tweets and their hashtags; the output is the dictionary $\mathscr{E}=\{(h,\bv_h) \,|\, \forall h\}$ containing all the hashtags and their embeddings expressed in the same semantic space as the tweet embeddings.}
\label{fig:BERTag}
\end{figure}
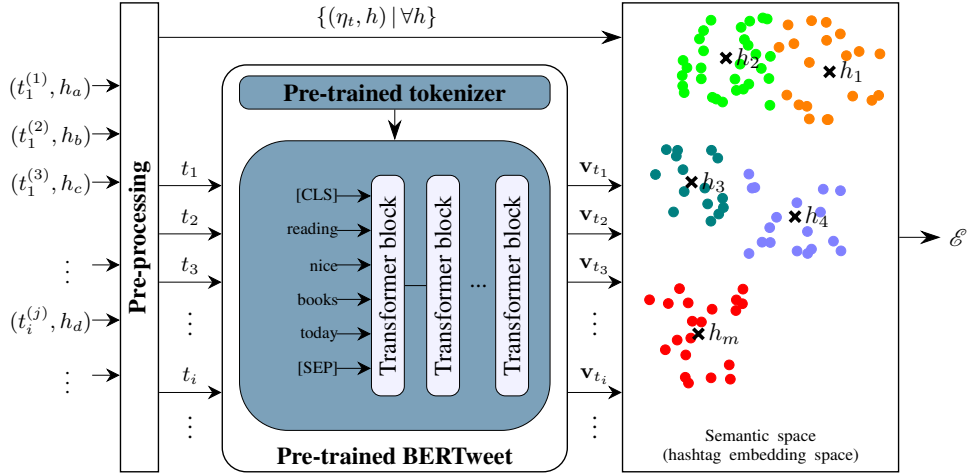

The architecture of BERTag comprising these three stages is shown in Fig.~\ref{fig:BERTag}.

\subsubsection{Pre-processing}
\label{sec:Pre-processing}
The pre-processing stage of the BERTag module consists of two steps.

The first step is to perform data cleaning on the tweets. This step involves removing stop words, mentions, URLs, and punctuation except for the \# sign. Tweets are transformed into lower case letters and non-English tweets are removed.

The second step is to load the pre-trained BERTweet models from Hugging Face\footnote[1]{https://huggingface.co/vinai/bertweet-base}. These pre-trained models contains at least 16 billion word tokens generated via the TweetTokenizer from NLTK\footnote[2]{https://www.nltk.org}. The tokenization process removes all the \textit{mentions} from tweets and truncates the length of repeated characters to 3 to reduce the number of unique words (e.g., `Heeeeelllllllo' becomes `Heeelllo'). TweetTokenizer keeps hashtags as they are and considers them as independent tokens. 
    We use two versions of BERTweet in the evaluation of our framework:
    \begin{itemize}
        \item BERTweet-base, and
        \item BERTweet-covid19-base-uncased.
    \end{itemize}
    Both versions of BERTweet accept any tweet in the following form:
    The tweet has been tokenized; the `[CLS]' token has been appended to the start of the tweet;  the `[SEP]' token has been appended at the end of the tweet; each token for the tweet has been mapped to a unique token ID; the tweet has been truncated to a pre-defined length; and attention masks for the `[PAD]' tokens have been added to differentiate between padding and non-padding.

\subsubsection{Extracting tweet embeddings}
At the end of the second pre-processing step above, each tweet is represented by a BERTweet-base or a BERTweet-covid19-base-uncased embedding vector. Following the dimensions specified in BERTweet, the tweet embedding vectors in BERTag are also 768 dimensional. For any given tweet $t$, we use $\bv_t$ to denote its tweet embedding vector.

Since the tweet embeddings in BERTag are learned from a large corpus of tweets, tweets with similar textual contents are closer together in the semantic space. The  cosine distance function (i.e., $1-\cos(\theta)$, where $\theta$ is the angle between the two embedding vectors being compared) can therefore be used as a distance measure between tweets.

\subsubsection{Computing hashtag embeddings}

Since every tweet $t$ is now paired with a single hashtag label $h$, through the index $\{(\eta_t,h) \,|\, \forall t\}$, all tweets associated with the hashtag $h$ can be retrieved. As described in the previous subsection, the way that the BERTag tweet embeddings are learned ensures that tweets which share the same hashtag are near each other and form a cluster, because these tweets have similar textual contents. Thus, the number of clusters can be set to $m$, the number of unique hashtags in the dataset.

Let $\{ \bv_t \,|\, \forall t \}$ be the set of embedding vectors of tweets sharing $h$ as a common hashtag, the embedding $\bv_{h}$ for hashtag $h$ can be easily calculated by the arithmetic mean: 
$\bv_{h}= \frac{1}{s} \sum_{t} \bv_{t} / n_h$,
where $n_{h}$ is the total number of tweets that use hashtag $h$ and $s$ is a normalisation term to ensure that $\bv_h$ is a unit vector. The centroid of each cluster thus represents a hashtag embedding in the semantic space alongside its tweets. 
Putting all the embedding vectors $\bv_{h}, \forall h$ together yields a dictionary $\mathscr{E}$ of hashtags with their embeddings, i.e., $\mathscr{E}=\{(h,\bv_{h}) \,|\, \forall h\}$.
From hereon, we refer to the hashtag embeddings computed from BERTweet-based and BERTweet-covid-base-uncased (see Section~\ref{sec:Pre-processing}) as \textit{BERTag-base} and \textit{BERTag-covid}.

\subsubsection{BERTag hashtag updates}

Our BERTag module considers the dynamic nature of social media platforms where new tweets and hashtags are continuously posted over time. So, it is designed to avoid recalculating the hashtags' embeddings from scratch when new tweets are added. It is straightforward to incorporate a new tweet embedding and calculate the cumulative average of all the tweet embeddings to get the resultant hashtag embedding. Let $n_h$ be the number of existing tweets that have been found to use hashtag $h$ and ${\bf v}_h$ be the corresponding hashtag embedding. Let ${\bf v}_t,$ be the embedding vector of the new tweet $t$. Then the inclusion of this new tweet $t$ results in the following updates:
\begin{align}
{\bf v}_h & \gets \frac{1}{s} \left(\dfrac{n_h}{n_h+1} \bv_{h} + \dfrac{ 1}{n_h+1} \bv_{t}\right), \label{eq:centroid2} \\
n_h & \gets n_h + 1,
\end{align}
where $s$ is a scale factor to normalise $\bv_h$ to a unit vector.

\subsection{Synonym and Thesaurus Construction}
\label{sec:knn}
The process of building a thesaurus starts with identifying the synonyms of each existing hashtag. For every hashtag $h \in H$, the set of its $k$ closest synonyms are identified from $\mathscr{E}=\{(h,\bv_{h})\}$ using the $k$-nearest neighbours (kNN) algorithm. Here, the value of $k$ is the number of desired synonyms. As kNN includes the hashtag $h$ itself as one of its own synonyms, to get, for instance, $n$ synonyms for $h$, we set the $k$ value in kNN to $n+1$. We use the cosine distance as a measure of closeness of hashtags in the hashtag embedding space. This process gives a list of synonyms $\syn(h)$ for each hashtag $h$. These lists of synonyms together form the thesaurus $\thes$. That is, $\thes = \{(h, \syn(h)) \,|\, \forall h\}$. Algorithm~\ref{alg:syn_construction} shows the pseudocode for constructing the list of $k$ synonyms for a given hashtag.

\begin{algorithm}[tbp!] \small			
	\caption{~~ Construct\_Synonyms }
	\label{alg:lookup}
	\begin{flushleft}
		\textbf{INPUT:} \\
        \hspace*{\algorithmicindent} $\hat{h}$: a recommended hashtag.  \\
        \hspace*{\algorithmicindent} $k$: an integer denoting the number of desired synonyms.  \\
        \hspace*{\algorithmicindent} $\mathscr{E}=\{(h,\bv_{h})\}$: the dictionary from BERTag training. \\
		\textbf{OUTPUT:} $\hat{S}$: the set of synonyms for $\hat{h}$. \\
	\end{flushleft}	
	\begin{algorithmic}[1]
        \STATE \LCOMMENT{Get the embedding for $\hat{h}$}
        \FOR {each hashtag $h \in \mathscr{E}$} 
            \IF {$\hat{h}$ matches $h$}
                \STATE $\bv_{\hat{h}} \gets \bv_h$ 
                \STATE break
            \ENDIF
        \ENDFOR
        \STATE Construct $S_v$ using kNN and $\mathscr{E}$ to find $k+1$ hashtags having shortest cosine distances from $\bv_{\hat{h}}$
        \STATE Construct $\hat{S}$ by searching $\mathscr{E}$ for all hashtags in $S_v$
        \STATE return $\hat{S}$
	\end{algorithmic}
	\label{alg:syn_construction}
\end{algorithm}

Throughout the paper, we use $\syn(h)$ to denote the set of synonyms with $h$ itself included. Where appropriate, we use a subscript to denote the number of synonyms (other than the hashtag itself) in the set. For instance, 
\begin{enumerate}[label={\footnotesize{$\bullet$}},itemsep=0ex]
\item $\syn_0(h) = \{h\}$;
\item $\syn_4(h)$ is a list containing five hashtags, including $h$. 
\end{enumerate}
When the argument to $\syn()$ is a set of hashtags, e.g., if $S=\{h_1, h_2, \cdots \}$, then $\syn(S) \triangleq \underset{h_i\in S}{\cup} \syn(h_i)$.

It should be noted that the synonym relationship between hashtags is not symmetrical. In other words, suppose that a hashtag $h_i$ is in the synonym set of another hashtag $h_j$, then it is not necessary that $h_j$ is in the synonym set of $h_i$. For instance, using the example shown in Table~\ref{tab:examples}, \#sport is one of the three closest synonyms of \#swim; however, \#swim is outside the three closest synonyms of \#sport.

\subsection{Semantic Evaluation}
\label{sec:sematic-eval}
The third and last module of the \#REval framework is the \hrRE\ calculation.
\subsubsection{$\hrRE$}
\label{sec:hrRE}

Let $R$ and $G$ be, respectively, the sets of recommended hashtags and ground truth hashtags. In our previous work~\cite{alsini2020hit}, the hit ratio is defined as:
\begin{equation}
\hr(R,G)= \frac{|R \cap G|}{\min(|R|,|G|)},
\label{eq:hr}
\end{equation}
where $|\!\cdot\!|$ denotes the cardinality of the set.
In the \#REval framework, we modify this formula to incorporate the sets of synonyms extracted from $R$ as follows:
\begin{flalign}
&\hrRE(R,G) =  \nonumber \\
& \frac{1}{\min(|R|,|G|)} \begin{cases}
  \sum\limits_{\hat{h}\in R} \rho\left(\syn(\hat{h}) \cap G \neq \emptyset\right) & \text{if } |R| \le |G| \\
  \sum\limits_{h\in G} \rho\bigg(h \in \syn(R)\bigg) & \text{otherwise},
\end{cases}
\label{eq:hrREval}
\end{flalign}
where 
\[
\rho(s)=
\begin{cases}
  1 & \text{if $s$ = True}\\
  0 & \text{otherwise}.
\end{cases}
\]
We can see the original formula for hit ratio in Eq.~\eqref{eq:hr} is just a special case of the formula in Eq.~\eqref{eq:hrREval} with the number of synonyms being zero. 

Notice that synonyms are only constructed for the set of recommended hashtags $R$, and not for the set of ground truth hashtags $G$. By doing so, we consider each recommended hashtag $\hat{h}$ to be a \textit{hit} only if a ground truth hashtag $h$ in $G$ is synonymous to it, but not the other way round. Consider the case where the recommended hashtag is a more specific term but the ground truth hashtag is a more general term, such as  $R = \{\text{\#hockey}\}$ and $G=\{\text{\#sport}\}$. In this case, \textit{hockey} is a kind of \textit{sport} and it is more likely that $\syn(\text{\#hockey}) \ni \text{\#sport}$  and the \hrRE\ formula will correctly count the recommendation as a hit. On the other hand,  if $R= \{\text{\#sport}\}$ and $G=\{\text{\#hockey}\}$, then it is unlikely that $\syn(\text{\#sport}) \ni \text{\#hockey}$ as \textit{sport} is not necessarily about \textit{hockey}. In this case, the formula will correctly not count the recommendation as a hit. In contrary, if the synonym lists were allowed to grow for both the recommendation set $R$ and ground truth set $G$, then for extremely large $k$ values, the two synonym lists will overlap even though the starting hashtags in the two sets are very different. By restricting $G$ to remain as is, the above scenario will not occur. It should be noted that $k$ is usually required to be a large number due to the noise in hashtags; however, one should set the value of $k$ sensibly. If we were to let $k\rightarrow\infty$, then every hashtag in the dataset would be in the list of synonyms of any given hashtag, and it would not be meaningful to use the measure for hashtag recommendation evaluation.

\subsubsection{Synonym matching}

The pseudo code for computing the number of matches between a set of recommended hashtags and a set of ground truth hashtags is given in Algorithms~\ref{alg:syn_matching}. It is easy to verify that $\rho$, the number of matches  between the two sets $R$ and $G$ computed using Eq.~\eqref{eq:hrREval}, satisfies the condition: $\rho \le \min(|R|,|G|)$. The \#REval-hit-ratio is then computed as $\rho/\min(|R|,|G|)$. So, like the hit ratio defined in Eq.~\eqref{eq:hr}, \#REval-hit-ratio values are also bounded between 0 and 1, and the higher they are (closer to 1), the better. 

\begin{algorithm}[tbp!] \small			
 	\caption{~~ Match\_Synonyms }
 	\label{alg:hit}
 	\begin{flushleft}
 		\textbf{INPUT:}  \\
         \hspace*{\algorithmicindent} $R$: a set containing the recommended hashtags,  \\
         \hspace*{\algorithmicindent} $G$: a set containing the groundtruth hashtags, \\
         \hspace*{\algorithmicindent} $k$: the number of desired synonyms, \\
         \hspace*{\algorithmicindent} $\thes = \{(h, \syn(h)) \,|\, \forall h \text{ in the test set}\}$: a thesaurus. \\
      
 		\textbf{OUTPUT:} ${\rho}$: the number of matches. \\
 	\end{flushleft}	
 	\begin{algorithmic}[1]
 	\STATE \textbf{Initialize:} $\rho \gets 0$\\
 	\IF {$|R| \le |G|$}
 	    \FOR {$\hat{h} \in R$}
 	        \STATE $\syn(\hat{h}) \gets$ Look up $\hat{h}$ in $\thes$
 	        \IF {$\syn(\hat{h}) \cap G \neq \emptyset$}
 	            \STATE $\rho \gets \rho+1$
 	        \ENDIF
 	    \ENDFOR
 	\ELSE
        \STATE \LCOMMENT{construct the list of synonyms for $R$} \\
        \STATE $\syn(R) \gets \emptyset$ \\
        \FOR {$\hat{h} \in R$}
            \STATE $\syn(\hat{h}) \gets$ Look up $\hat{h}$ in $\thes$
            \STATE $\syn(R) \gets \syn(R) \cup \syn(\hat{h})$
        \ENDFOR
        \FOR {$h \in G$}
            \IF {$h \in \syn(R)$}
                \STATE $\rho \gets \rho+1$
 	        \ENDIF
        \ENDFOR
    \ENDIF

    \STATE return $\rho$
    \end{algorithmic}
    \label{alg:syn_matching}
\end{algorithm}

Four small examples showing how synonym matching works are given in Table~\ref{tab:examples}, for different cardinalities of $R$ and $G$. The last column of the table contains the $\hrRE$ values, expressed as a fraction. Note that both the recommended and ground truth hashtags can contain spelling mistakes, such as $\keepfit$ shown in the examples. Except for the third example where $R = \{\hockey\}$, all the other three examples show that the recommended hashtags meaningfully match with the ground truth hashtags; however, an exact matching of the hashtags in $R$ and $G$ will give a hit-ratio of zero for all the examples.

\begin{table}[tbp!]
\caption{Four examples showing how $\rho$ and $\hrRE$ are computed for different sizes of $R$ and $G$.
 }
\vspace{-0.5ex}
\label{tab:examples}
\renewcommand{\arraystretch}{1.2}
\setlength{\tabcolsep}{1pt}
    \centering \small
    \resizebox{.65\columnwidth}{!}{
    \begin{tabular}{| L{4cm} | L{2.9cm} | C{0.9cm} | C{2cm} |}
       \hline
       \centering{$R$}  & \centering{$G$} & $\rho$ & \#REval-hit-ratio\\ \hline\hline
       $\{\hockey, \championship\}$ & $\{\football, \sport\}$  & 1 & 1/2 \\ \hline
       $\{\football, \sport\}$ & $\{\hockey, \sports\}$ & 1 & 1/2 \\ \hline
       $\{\hockey\}$ & $\{\football, \rugby\}$  & 0 & 0/1 \\ \hline
       $\{\swim,\exercise\}$ & $\{\sport\}$  & 1 & 1/1 \\ \hline\hline
       \multicolumn{4}{|l|}{\parbox[t]{9cm}{\small 
       $\syn(\hockey)\!=\!\{\hockey,\bowling,\golf,\sport\}$ \\ $\syn(\championship)\!=\!\{\championship,\champion,\winner,\tournament\}$ \\ 
       $\syn(\football)=\{\football,\soccer, \footy,\rugby\}$ \\
       $\syn(\sport)=\{\sport,\sports,\exercise,\keepfit\}$ \\
       $\syn(\swim)=\{\swim,\dive,\paddle,\sport\}$ \\
       $\syn(\exercise)=\{\exercise,\keepfit,\yoga,\walking\}$}} \\
       \hline
    \end{tabular}}
\end{table}

\smallskip

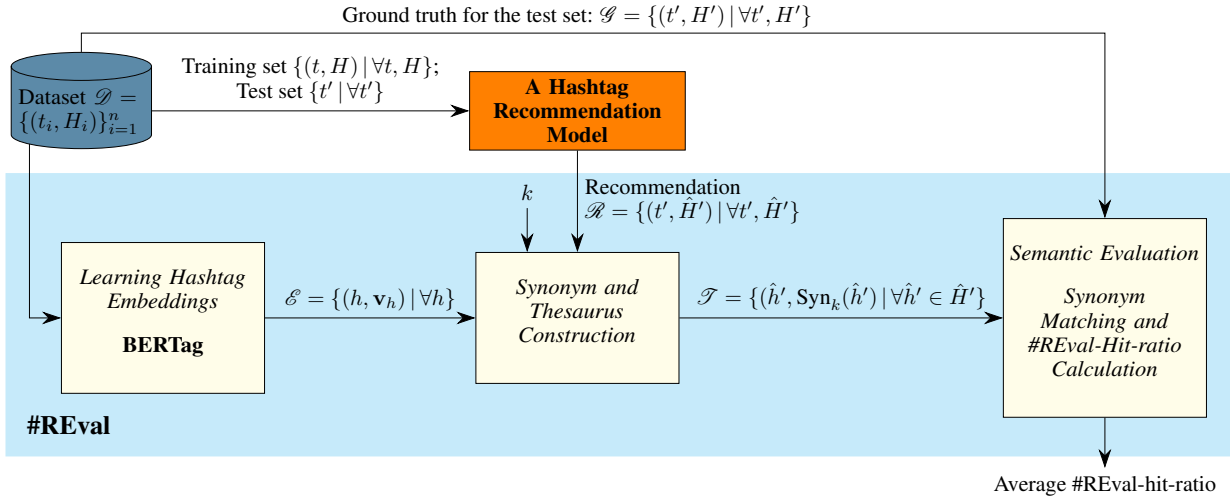
\begin{figure*}[tbp!]
\begin{center}
\resizebox{\textwidth}{!}{
    \begin{tikzpicture}
[REvalStyle/.style={rectangle, draw=black, fill=yellow!10, align=center},
HashtagModelStyle/.style={rectangle, draw=black, fill=orange, align=center},
blank/.style={rectangle, draw=none, align=center},
arrow/.style={-{Stealth[length=3mm, width=2mm]}}]

\filldraw [fill=cyan!20, draw=cyan!20] (11.5,4.5) rectangle (-8,0);
\node[rectangle, draw = none] at (-7,0.5){\large \bf \#REval};

\node [REvalStyle, text width=3cm] (REval1) at (-5.5,2.2) {
  \textit{\\[1ex] Learning Hashtag Embeddings} \\[2ex]  \textbf{BERTag} \\[1ex] ~};
\node[REvalStyle, text width=3cm] (REval2) at (1.1,2.2) {
  \textit{\\[1ex]Synonym and \\ Thesaurus \\  Construction} \\[1ex]~ };
\node[REvalStyle, text width=3cm] (REval3) at (9.5,2.2) {
  \textit{\\[1ex]Semantic Evaluation \\[2ex] Synonym \\ Matching and \\ \#REval-Hit-ratio \\ Calculation} \\[1ex]~};

\node[cylinder,
  draw = black,
  text = black,
  cylinder uses custom fill,
  cylinder body fill = airforceblue,
  cylinder end fill = airforceblue,
  aspect = 0.2,
  shape border rotate = 90, text width=2cm] (dataset) at (-6.8,5.5)
    {Dataset $\mathscr{D}=\{(t_i,H_i)\}_{i=1}^n$};

\node[HashtagModelStyle, text width=3.2cm] (HashtagAlg) at (1.1,5.5){
  \textbf{A Hashtag \\ Recommendation \\ Model}};

\draw [arrow] (REval1.east) -- (REval2.west)
  node[midway, above, blank] {$\mathscr{E}=\{(h,\mathbf{v}_h)\,|\,\forall h\}$};
\draw [arrow] (REval2.east) -- (REval3.west)
  node[midway, above, blank] {$\thes=\{(\hat{h}',\syn_k(\hat{h}')\,|\, \forall \hat{h}'\in\hat{H}'\}$};

\draw [arrow] (HashtagAlg.south) -- (REval2.north)
  node[midway, right, align=left] {Recommendation \\$\mathscr{R}=\{(t',\hat{H}')\,|\, \forall t',\hat{H}'\}$};

\draw [arrow] (REval3.south) -- ++(0, -0.8)
  node[below, blank] {Average $\hrRE$};
\draw [arrow] ( $(REval2.north) + (-0.8,0.7)$ ) -- ( $(REval2.north) + (-0.8,0)$ )
  node[above,yshift=20pt] {$k$};

\draw [arrow] (dataset.east) --  (HashtagAlg.west)
  node[midway, above, blank] {Training set $\{(t,H)\,|\,\forall t, H\}$; \\
  Test set $\{t'\,|\,\forall t'\}$} ;

\draw [arrow] (dataset.north) -- ++(0, 0.3)  -|  (REval3.north);

\draw [arrow] ( $(dataset.south west) - (0.25, -0.05)$)
  |- ( $(REval1.west) + (-0.2,0)$ ) -- (REval1.west);

\node[blank] (testing) at (1.1, 7.0) {Ground truth for the test set: $\mathscr{G}=\{(t',H')\,|\,\forall t', H'\}$};



\end{tikzpicture}
}
\end{center}
\vspace{-2ex}
\caption{The \#REval framework comprises three major modules: the \textit{BERTag} module, the \textit{Synonym and Thesaurus Construction} module, and the \textit{Semantic Evaluation} module.}
\label{fig:REval}
\end{figure*}

The \#REval framework comprising all the modules explained above is shown in Figure~\ref{fig:REval}.

\section{Experiments}
\label{sec:Experiments}

We have evaluated the effectiveness of our \#REval framework using three datasets, with the last dataset being domain-specific (in health). We have also tested the impact on the average \hrRE\ when embeddings other than BERTag-base and BERTag-covid are used in the first BERTag module of the framework.

\subsection{Datasets}
\label{Dataset}

\subsubsection{The AU Trends dataset} 

We used 16 Australian trending hashtags (see Table~\ref{tab:trends}) as seeds for crawling for tweets posted on March $13^{\text{th}}$, 2020. As other hashtags were also used in these tweets, the resultant number of unique hashtags is a lot larger than the number of trending hashtags. After the data cleaning step, 272,686 tweets containing one or more hashtags remain. The number of unique hashtags in these tweets is 15,375. As many hashtags were repetitively used multiple times, the total number of hashtags (including the repetitions) is 547,227. Table~\ref{tab:summary} summarises the statistics of this dataset and the remaining two datasets described below.

\begin{table}[tbp!]
\centering
\caption{Trending hashtags in Australia and US on March $13^{\text{th}}, 2020$}
\label{tab:trends}
\resizebox{.75\columnwidth}{!}{
\begin{tabular}{|@{\hskip3pt}l|@{\hskip3pt}l|}
\hline
\textbf{\centering AU Trends} & \textbf{\centering US Trends} \\ \hline\hline
\parbox[t]{3.5cm}{\#AustralianGP\\ \#Covid\_19\\ \#HeartbreakWeather\\ \#yiayhi\\ \#Formula1\\ \#SeizetheCSWmoment\\ \#australiangrandprix\\ \#WorldSleepDay\\ \#Ride2School\\ \#FlattenTheCurve\\ \#loona1stwin\\ \#SocialDistancing\\ \#MelbourneGP\\ \#ScottyFromMarketting\\ \#tomhanks\\ \#My750} & 
\parbox[t]{8cm}{\#HeartbreakWeather\\ \#FridayThe13th\\ \#QuarantineAndChill\\ \#BernieIsRight\\ \#panicbuying\\ \#WeGotThisSeattle\\ \#EverythingsGonnaBeOkay\\ \#selfisolating\\ \#MyQuirkiestQuirkReasonsToLeaveTheToiletSeatUp\\ \#Restore4GinKashmir\\ \#peteonkimmel\\ \#LUVVsTheWorld2\\ \#NiallCarpool\\ \#foreverfletcher} \\
\hline
\end{tabular}}
\end{table}

\begin{table}[tbp!]
\centering
\caption{Statistics summary of the four datasets used in our experiments.}
\label{tab:summary}

\resizebox{0.9\columnwidth}{!}{
\renewcommand{\arraystretch}{1.35}
\begin{tabular}{| L{.45\linewidth} | R{.16\linewidth} | R{.16\linewidth}  | R{.1\linewidth}  |}
\cline{2-4}
\multicolumn{1}{c|}{} & \textbf{AU Trends} (Mar 2020) & \textbf{US Trends} (Mar 2020) & \textbf{Health} (2015) \\ \hline
No.~of hashtagged tweets & 272,686 & 146,963 & 11,210 \\ \hline
No.~of unique hashtags & 15,375 & 5,608 & 2,710 \\ \hline
No.~of hashtags (including repetitions) & 547,227 & 206,663 & 15,548 \\ \hline
Min.~no.~of hashtags per tweet & 1 & 1 & 1 \\ \hline
Max.~no.~of hashtags per tweet & 14 & 13 & 8 \\ \hline
Avg.~no.~of hashtags per tweet & 2.1 & 1.3 & 1.2 \\ \hline
No.~of training tweets (after pre-processing) & 242,032 & 125,781 & 9,900 \\ \hline
No.~of testing tweets (after pre-processing) & 26,893 & 13,976 & 1,101 \\ \hline
\end{tabular}}
\end{table}

\subsubsection{The US Trends dataset} 

Following the same tweet crawling process above, we used 14 US trending hashtags (see Table~\ref{tab:trends}) to collect tweets, also on March $13^{\text{th}}$, 2020. 

\subsubsection{The Health dataset}

This dataset~\cite{Karami-2018Fuzzy} contains health-related tweets collected in 2015 from 16 news agencies such as BBC, CBC, CNN, FoxNews, NBC.

\subsection{Experimental setup}

\subsubsection{Variants of the \#REval framework}
As shown in Fig.~\ref{fig:REval}, the \textit{BERTag} module (see Section~\ref{sec:bertag}) outputs BERTag-base embeddings or BERTag-covid embeddings for the dictionary $\mathscr{E}$. Apart from using these embeddings, a different way to learn the hashtag embeddings for building the dictionary $\mathscr{E}$ can be evaluated. In our experiments, we tested the Word2Vec and FastText models.  The training procedure is summarised below.

We firstly trained Word2Vec (similarly for FastText) model using the \textit{gensim} library\footnote[3]{https://radimrehurek.com/gensim/index.html}, where every tweet is considered as a document. Knowing that tweets are short texts, we set the \textit{window\_size} hyperparameter to 2 to determine the dependency of a word to its neighbouring words. We took into account every word in the tweets because some hashtags occur only once. Thus, we set the \textit{min\_count} hyperparameter to 1. The number of training epochs was set to 30. Finally, a dictionary $\mathscr{E}=\{(h,\bv_{h})\}$ containing all the hashtags and their embeddings was generated. The hashtag embeddings were finally obtained by taking the arithmetic means of the tweet embeddings obtained from the training process.  

So, in our experiments, we compared four different variants for the first module of the \#REval framework shown in Fig.~\ref{fig:REval}. Each version uses:
\begin{itemize}
    \item hashtag embeddings defined in terms of Word2Vec;
    \item hashtag embeddings defined in terms of FastText;
    \item BERTag-base embeddings; or
    \item BERTag-covid embeddings.
\end{itemize}

\subsubsection{Generation of $\mathscr{R}$ for the test set}
\label{HR_model}
As shown in Fig.~\ref{fig:REval}, \#REval is agnostic to the hashtag recommendation model that produces the recommendation set $\mathscr{R}$. This is evident from the figure that the orange rectangular box  is outside the light blue shaded region for \#REval. However, in order to test the \textit{Semantic Evaluation} module of \#RERval, we need an example hashtag recommendation model. In this paper, we chose the hashtag recommendation method based on tweet similarity proposed in~\cite{Alsini-2020} to generate $\mathscr{R}$. This method encoded the tweets as \textit{Mean of Word Embedding} (MOWE) vectors computed from words that were trained using the Word2Vec model. For a given test tweet, the algorithm computed its MOWE vector and compared it with the training tweets' MOWE vectors based on their cosine similarity measure. A threshold value of $0.5$ was used as the cut off for selecting similar tweets from the training set. Hashtags that were used in these selected training tweets were ranked based on their \textit{hashtag popularity}, and the top-$r$ hashtags, for a pre-defined positive integer $r$, were put forwarded as the recommended hashtags for the test tweet. Because of the threshold value ($0.5$) used above, the number of recommended hashtags for any test tweet may be fewer than $r$. 

\subsubsection{Top-$r$ recommendations and number of synonyms}
For each dataset mentioned above, only those tweets having at least one hashtag are retained. After the pre-processing stage (see Section~\ref{sec:Pre-processing}), the tweets are split into 90/10\% to form the training set and the test set. 
The training set is used to train the hashtag recommendation method~\cite{Alsini-2020} described above to generate $\mathscr{R}$ for the test set. For each tweet in the testing set, the top-$r$ recommended hashtags are extracted for the following $r$ values: $1$, $5$, and $10$. 
This gives three separate $\mathscr{R}$ for the three top-$r$ recommendations for comparison with the ground truth test set $\mathscr{G}$ (see Fig.~\ref{fig:REval}).

For each recommended hashtag, we set the number of synonyms (the value of $k$ in Algorithms~\ref{alg:syn_construction} and~\ref{alg:syn_matching}) to a range of values: $\{0, 5, 10, 20, \cdots, 70\}$. The reason for setting $k$ to a large value such as $70$ is because hashtags are noisy text strings and can contain spelling mistakes, numbers, etc. For instance, the hashtags \#covid, \#covid0, \#covid19, \#covid2019, \#coronavirus, and \#coronav are all synonymous words and most of them are not in the standard English dictionary. So, when considering a specific target hashtag, we need to enlarge the number of synonyms in the hashtag embedding space. It should be noted that, for the case where $k=0$, the \#REval-hit-ratio becomes the hit-ratio (Eq.~\eqref{eq:hr}) and it can be considered to be the output from the baseline exact-matching method of hashtags.

\subsection{Results and Discussions}
\label{sec:Results_and_Discussions}

\begin{figure*}[t]
\begin{center}
   \subfloat{\includegraphics[height=0.6\textwidth]{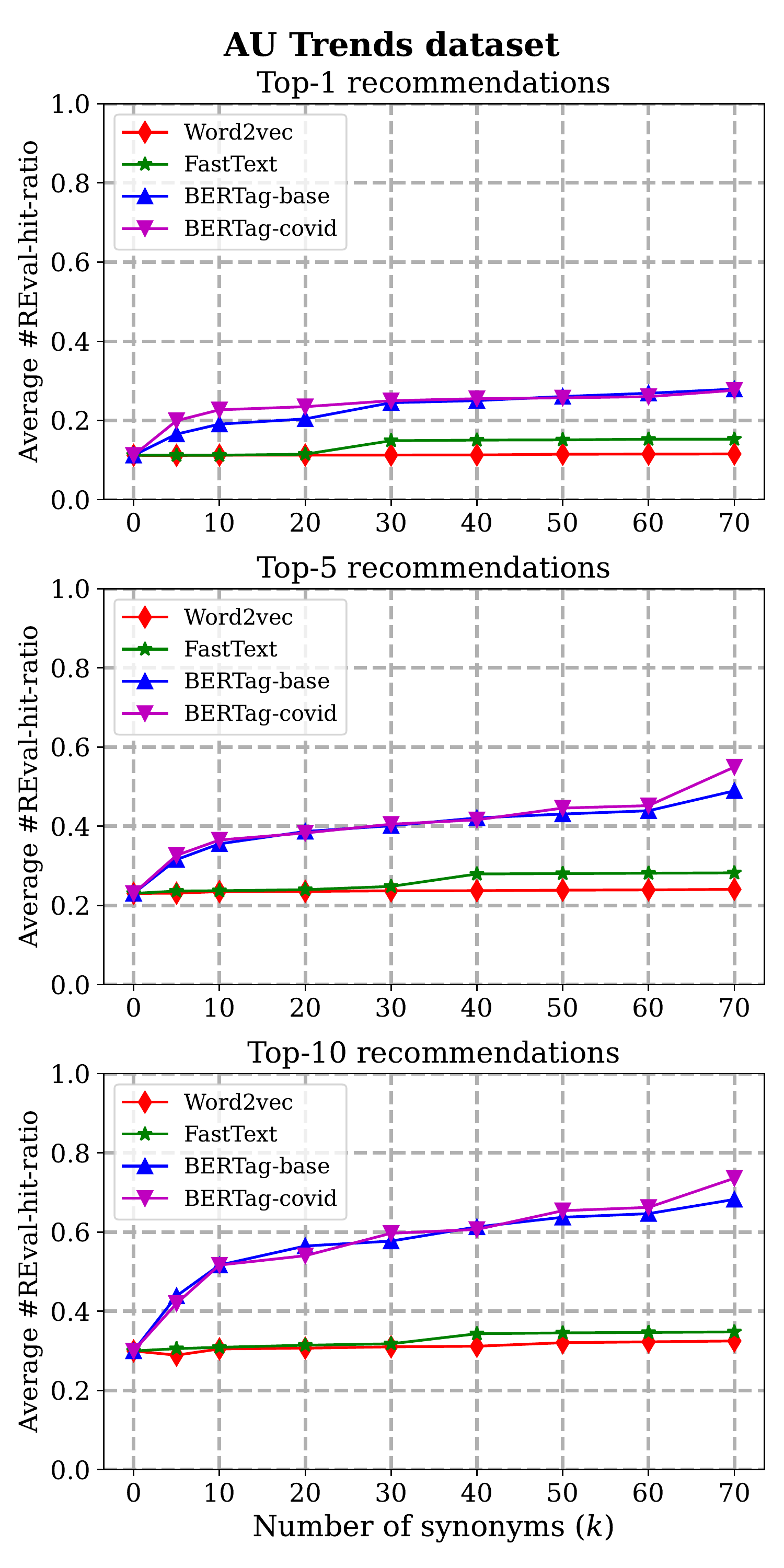}}
   \subfloat{\includegraphics[height=0.6\textwidth]{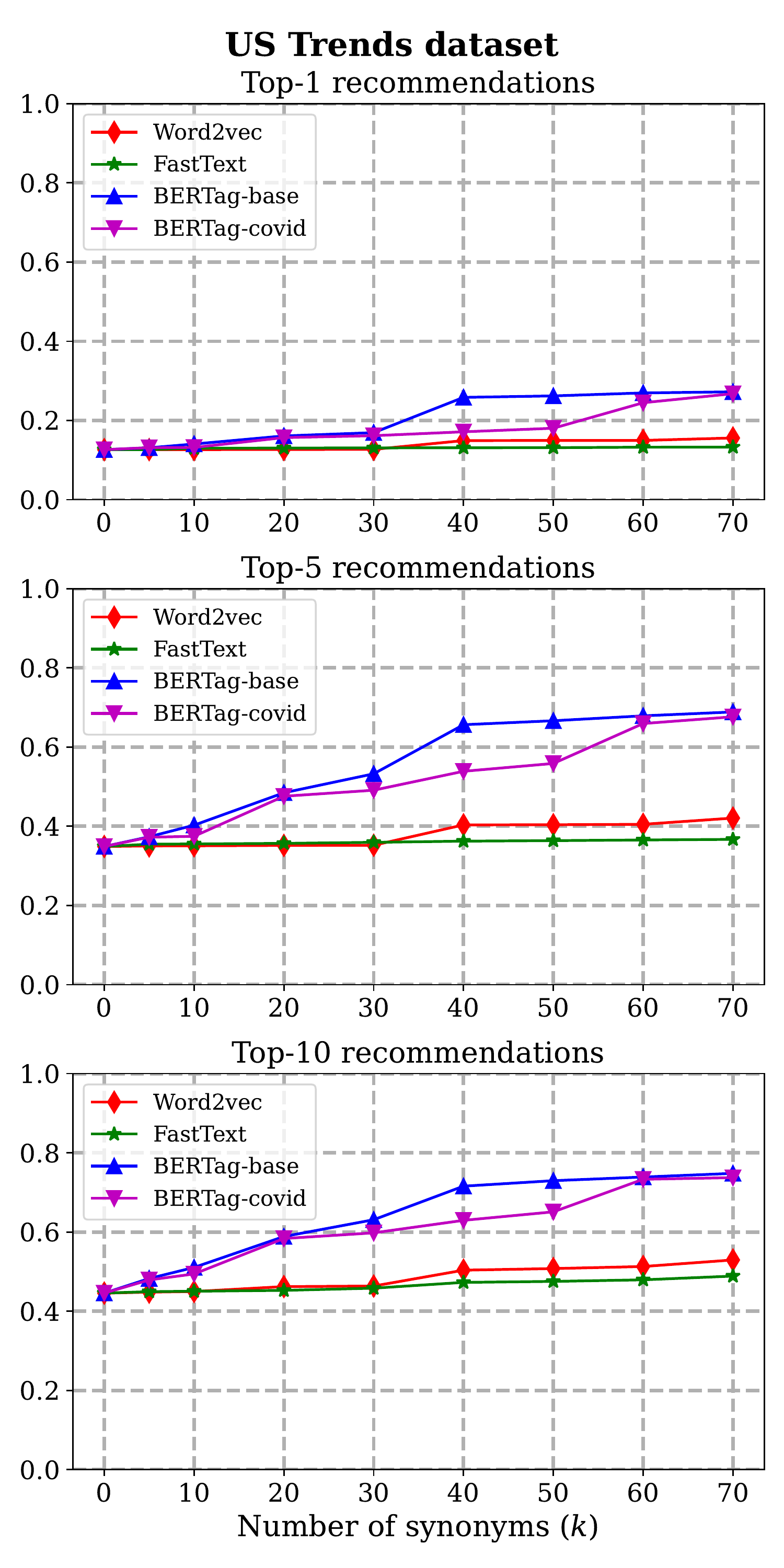}}
   \subfloat{\includegraphics[height=0.6\textwidth]{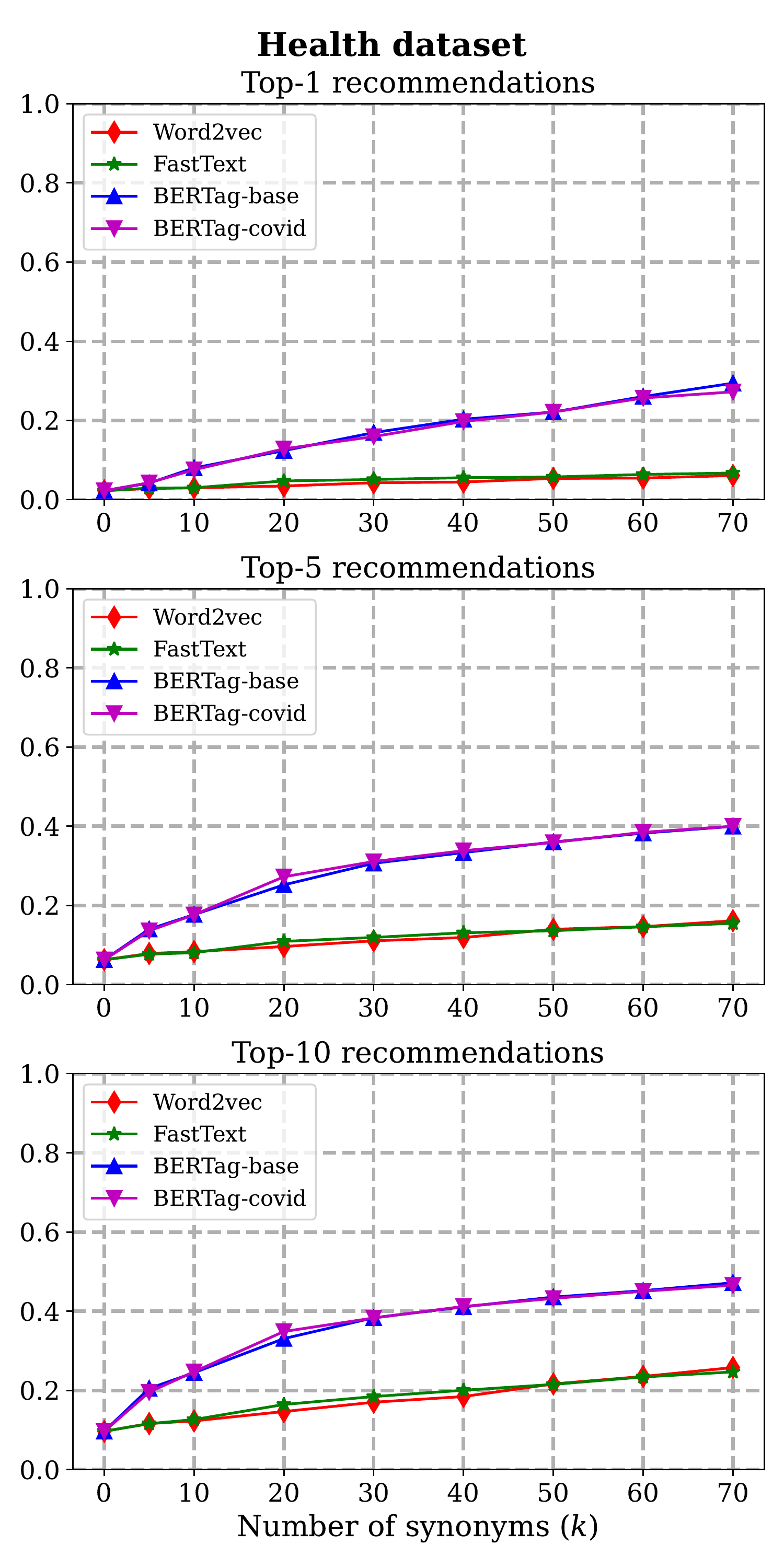}}
\end{center}
\caption{A comparison of the average \#REval-hit-ratios for top-$1$, top-$5$, and top-$10$ recommendations from the hashtag recommendation model of~\cite{Alsini-2020} on the three datasets, with the hashtag embeddings represented as Word2Vec, FastText, BERTag-base, and BERTag-covid. The number of synonyms ($k$) varies from $0, 5, 10, 20, \cdots, 70$. (figure best viewed in colour)}
\label{fig:performance_hit_ratio_BERTag}
\end{figure*}

For each test tweet $t'$ in the test set, we computed the \hrRE\ between the recommended hashtag set $R$ and the ground truth hashtag set $G$ for $t'$. The average \hrRE\ was finally computed for the entire recommendation set $\mathscr{R}$.  
Figure~\ref{fig:performance_hit_ratio_BERTag} shows the average \#REval-hit-ratios  for the four variants of embeddings computed using different numbers of synonyms (the value of $k$ on the horizontal axis) for the three datasets. The top, middle, and bottom rows of the figure correspond to the top-1, top-5, and top-10 recommendations.  The $\hrRE$ at $k=0$ in each subplot denotes the case where no synonyms were used, i.e., $\syn_0(h) = \{h\}$ for each hashtag $h$, and effectively the hit-ratio (Eq.~\eqref{eq:hr}) for the baseline exact-matching of hashtags was computed instead. 

Comparing the results shown in Figure~\ref{fig:performance_hit_ratio_BERTag}, we summarize the following four observations.
\smallskip

\noindent \textbf{\small (i)~The exact-matching method underrates the performance of the hashtag recommendation model.}
As expected, the baseline exact-matching method (where $k=0$) has the lowest average \hrRE\, in each subplot, as no synonyms were included in the calculation. For the AU Trends and Health datasets (Fig.~\ref{fig:performance_hit_ratio_BERTag}, columns 1 and 3) we can see that, by increasing $k$ slightly to 5 or 10, the average \hrRE\ values are almost double when the BERTag-base (blue curves) or BERTag-covid (magenta curves) embeddings were used to represent the hashtags. This improvement is consistent for all the top-1, top-5, and top-10 recommendations. For the US Trends dataset, we also observe increases in the average \hrRE\, values when $k$ grows, although the increases are smaller when $k$ is under 10. When Word2Vec or FastText was used as hashtag embeddings, the increase in average \hrRE\ with respect to $k$ is small.

Comparing across the three datasets, the hashtag recommendation model~\cite{Alsini-2017} performed the best on the US Trends dataset. Even when $k=0$, its  average \hrRE\ on the US Trends dataset is much higher than those for the other two datasets.  It should be noted that hashtag recommendation is not an easy task because of noise in the data and large number of hashtags and variations of tweet contents.  With synonyms included, its average \hrRE\ is just under 0.8 for the top-10 recommendations.

\smallskip

\noindent \textbf{\small (ii)~BERTag-base and BERTag-covid outperform Word2Vec and FastText.} Since the same recommendation test set $\mathscr{R}$ was used for all the four variants, the only difference is what embeddings were used to represent the hashtags and, consequently, the quality of the generated synonyms for the \hrRE\ calculation. We manually inspected many hashtags and their synonyms extracted from the embedding spaces of these four variants and we summarise six hashtags (two from each dataset) and their five closest synonyms in Table~\ref{tab:5_synonyms}. We can see from the table that the synonyms from BERTag-base and BERTag-covid are very similar in most examples. Also, compared to those from Word2Vec and FastText, these synonyms are more related to the original hashtags in the first column. For the hashtag \#stockmarket, both BERTag-base and BERTag-covid even picked up the names of two specific shares, \#sensex and \#nifty, as synonyms. Looking at the Word2Vec column in the table,  we see some strange words, such as \#lasagna and \#grilling as the two closest synonyms of \#aging; and \#ask as the first synonym of \#quinoa. FastText, on the other hand, seemed to group words that have similar substrings together in the hashtag space, e.g., the substring ``market'' appears in all the five synonyms of \#stockmarket; the postfix ``ing'' appears in four of the five synonyms of \#aging.  While these synonymous hashtags from FastText are sometimes quite relevant, there are many cases where FastText did not produce good synonyms, e.g., the five synonyms from FastText for \#quinoa are quite poor.

\begin{table*}[tbp!]
\caption{The five closest synonyms of a few example hashtags from the three datasets.}
\label{tab:5_synonyms}
\renewcommand{\arraystretch}{1.3}
\scalebox{0.73}{
\begin{tabular}{|c|l|l|l|l|}
\hline
 \multirow{2}{*}{\textbf{Hashtag}} & \multicolumn{4}{c|}{\textbf{Synonyms from thesaurus using}} \\ \cline{2-5} 
  & \multicolumn{1}{c|}{\textbf{Word2Vec}}  & \multicolumn{1}{c|}{\textbf{FastText}}  & \multicolumn{1}{c|}{\textbf{BERTag-base}}
  & \multicolumn{1}{c|}{\textbf{BERTag-covid}}  \\ \hline\hline
 
  \myCparbox{2.3cm}{\#stockmarket \\(AU Trends)}  &             
  \myparbox{4cm}{\#fearnoevil, \#mkonoh, \#awareness, \#sooryavanshion24thmarch, \#faculty} & 
  \myparbox{5cm}{\#stockmarkets, \#stockmarketcrash, \#stockmarket2020, \#trumpstockmarket, \#stockmarketcrash2020} &
  \myparbox{4.5cm}{\#stockmarketcrash, \#stockmarketcrash2020, \#sensex, \#nifty, \#stocks}          &
  \myparbox{4.5cm}{\#stockmarketcrash2020, \#stockmarketcrash, \#sensex, \#nifty, \#commodities}              \\ \hline
  \myCparbox{2.3cm}{\#selfisolation \\(AU Trends)}  &             
  \myparbox{4cm}{\#onthepulse, \#coronahitsnoida, \#koreandrama, \#julialang,  \#mohfw} & 
  \myparbox{5cm}{\#selfisolationhelp, \#selfisolating,  \#selfisolate, \#italyselfisolation, \#selfisolationlevel100} &
  \myparbox{4.5cm}{\#socialdistancing, \#selfquarantine, \#selfisolate, \#quarantine, \#isolation}          &
  \myparbox{4.5cm}{\#socialdistancing, \#selfquarantine, \#selfisolate, \#quarantine, \#isolation}              \\ \hline
 
  \myCparbox{2.3cm}{\#panicbuying \\(US Trends)}  & 
  \myparbox{4cm}{\#aiea, \#petebuttigieg, \#powernap, \#florida, \#ebooks}                &
  \myparbox{5cm}{\#panicbuyinguk, \#panicbuy, \#stoppanicbuying, \#panickbuying, \#costcopanicbuying}       &
  \myparbox{4.5cm}{\#cronavirus, \#selfisolating, \#blessed, \#superstore, \#foodbank}                         &
  \myparbox{4.5cm}{\#hoarding, \#cronavirus, \#panickbuying, \#superstore, \#asda}                         \\ \hline 
  \myCparbox{2.3cm}{\#selfisolation \\ (US Trends)}                                          & 
  \myparbox{4cm}{\#lka, \#stop, \#cinderella,\\  \#gaslightingofamerica,\\  \#melekoout}    & 
  \myparbox{5cm}{\#selfisolationhelp, \#selfisolating, \\ \#selfisolate, \#selfisolationhelpdrogheda,\\  \#isolation}      & 
  \myparbox{4.5cm}{\#selfquarantine, \#selfisolationhelp,\\  \#socialdistancing, \#lockdown,  \#quarantine}  & 
  \myparbox{4.5cm}{\#socialdistancing, \#selfquarantine, \\ \#selfisolationhelp,  \#selfisolating, \#quarantined} \\ \hline
  \myCparbox{2.3cm}{\#aging \\ (Health)}                                          & 
  \myparbox{4cm}{\#lasagna, \#grilling, \#healthforall, \#healthyliving, \#pet}    & 
  \myparbox{5cm}{\#antiaging, \#healthyaging, \#agility, \#jogging, \#breaking}      & 
  \myparbox{4.5cm}{\#memory, \#dementia, \#stroke, \#autism, \#health}  & 
  \myparbox{4.5cm}{\#memory, \#stroke, \#autism, \#dementia, \#health} \\ \hline
  \myCparbox{2.3cm}{\#quinoa \\ (Health)}                                          & 
  \myparbox{4cm}{\#ask, \#polio, \#hulahooping, \#brunch, \#cancerfighting}    & 
  \myparbox{5cm}{\#quit, \#quiz, \#quesadillas, \#mojo, \#smokers}       & 
  \myparbox{4.5cm}{\#healthy, \#recipes, \#health, \#diet, \#food}  & 
  \myparbox{4.5cm}{\#healthy, \#recipes, \#health, \#diet, \#food} \\ \hline
\end{tabular}
}
\end{table*}

\medskip

\noindent \textbf{\small (iii)~Both BERTag-base and BERTag-covid have similar performance and are better embeddings for hashtags.} Comparing the \hrRE s when these embeddings are used in the BERTag module (Fig.~\ref{fig:REval}), we see that BERTag-base (blue curve) outperforms BERTag-covid (magenta curve) slightly for the US Trends dataset, but the results are the other way round for the AU Trends dataset. For the Health dataset, the two curves almost completely overlap, except when $k=20$, the magenta curve is a little bit higher.  

\begin{figure*}[thbp]
\centering
\parbox{0.33\textwidth}{\centering\footnotesize\textbf{t-SNE projection of 85646 labelled tweets}}
\parbox{0.33\textwidth}{\centering\footnotesize\textbf{t-SNE projection of 52929 labelled tweets}}
\parbox{0.33\textwidth}{\centering\footnotesize\textbf{t-SNE projection of 3510 labelled tweets}}
\\[-2ex]
\subfloat[AU Trends dataset]{
  \includegraphics[width=.20\textwidth]{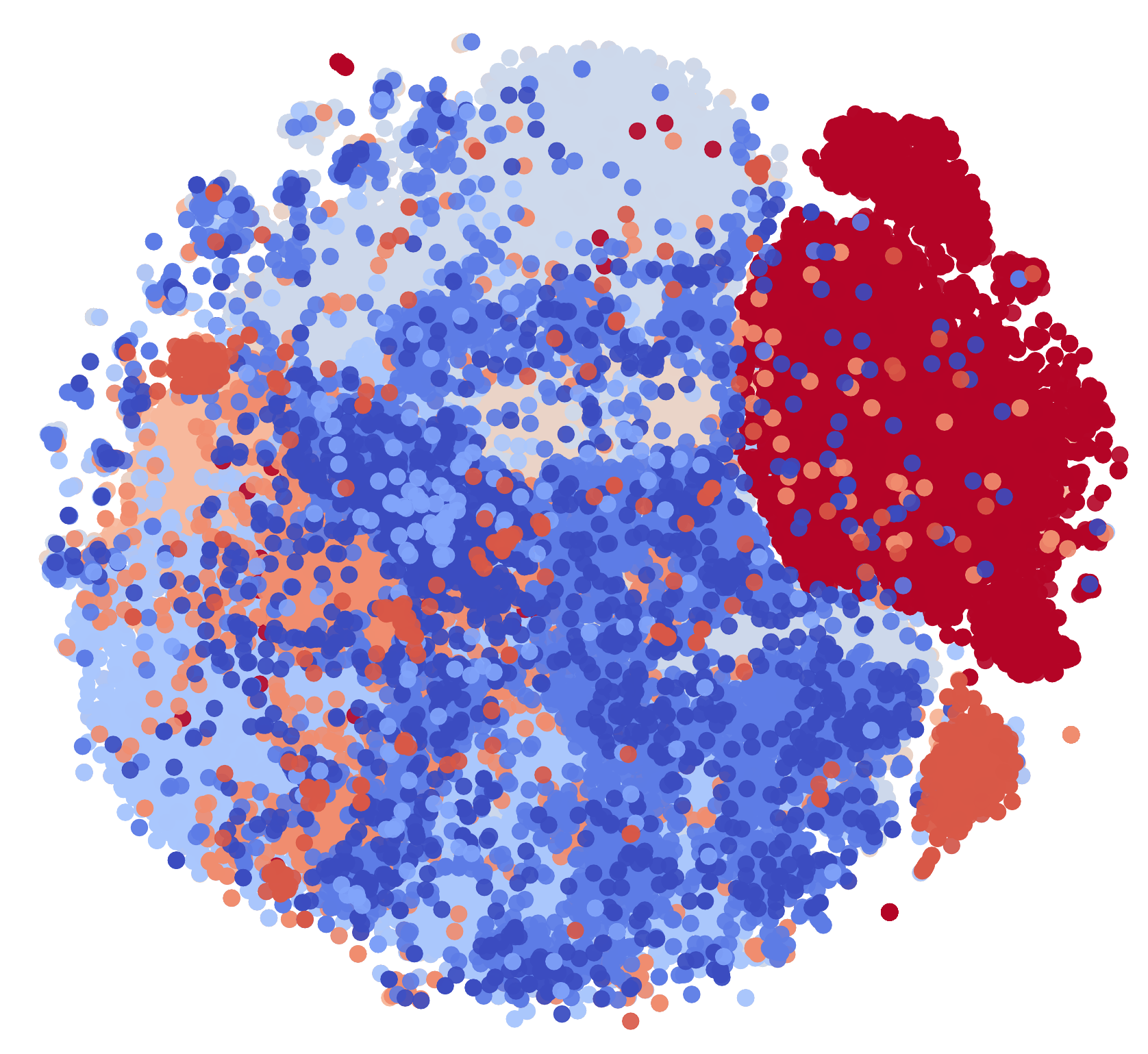}
  \includegraphics[height=.13\textwidth]{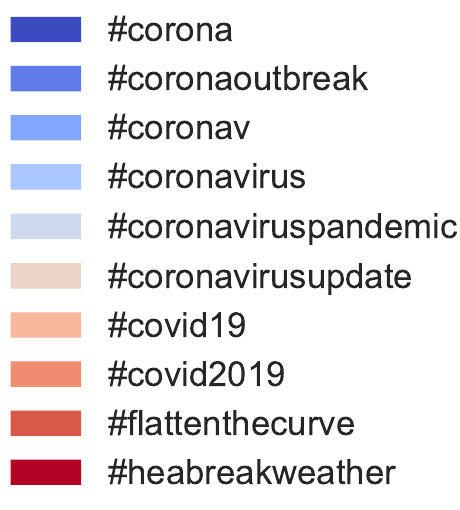}}
\subfloat[US Trends dataset]{
  \includegraphics[height=.20\textwidth]{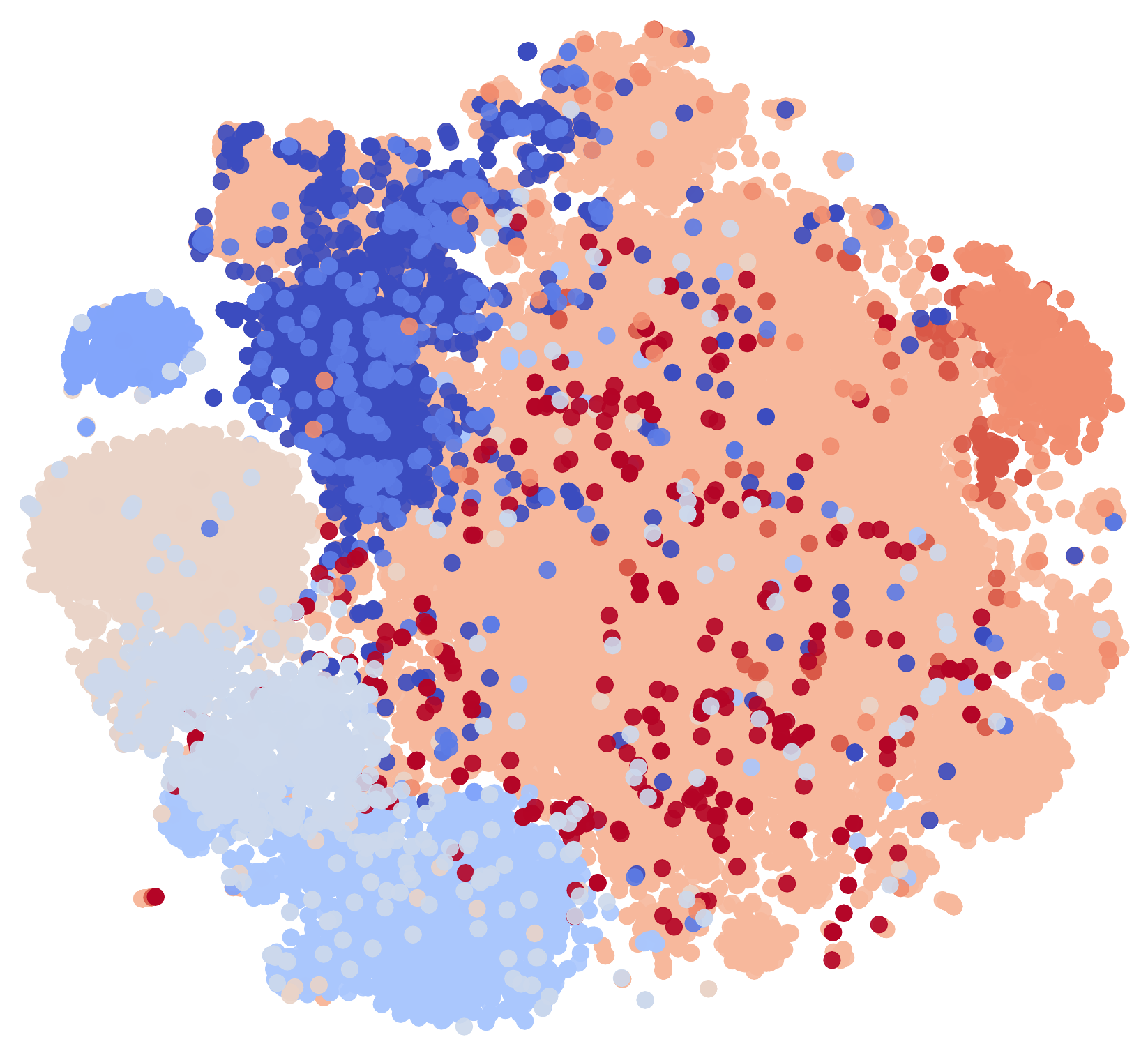}
  \includegraphics[height=.13\textwidth]{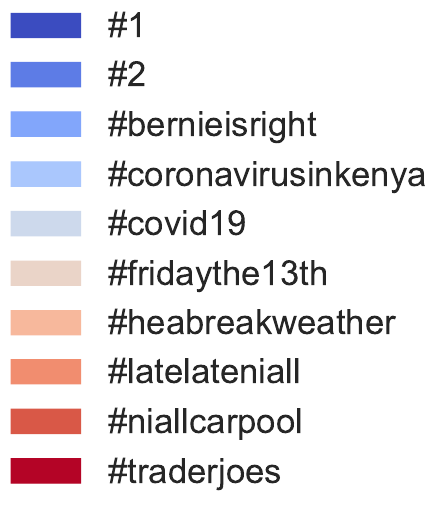}}
\subfloat[Health dataset]{
  \includegraphics[height=.20\textwidth]{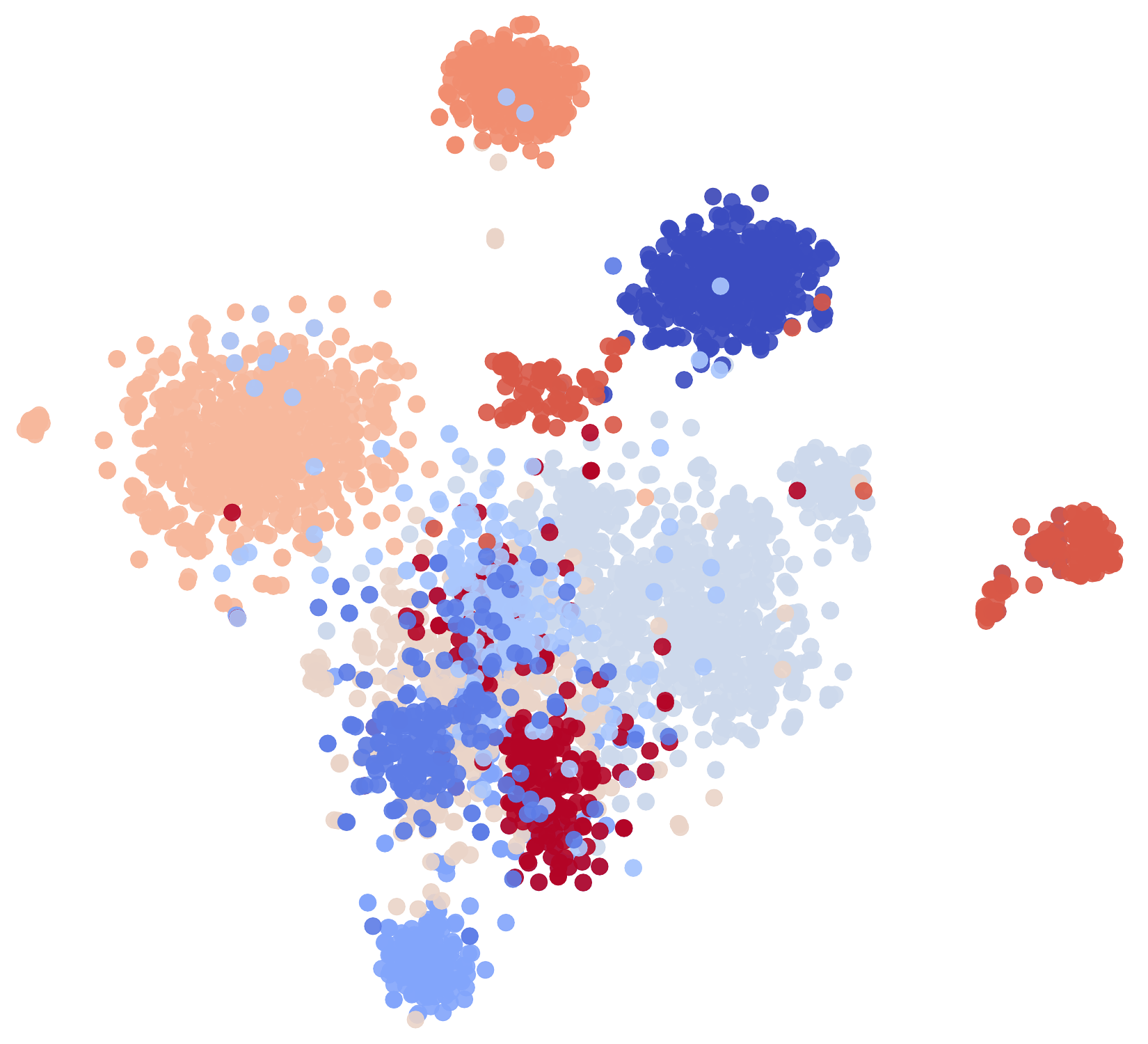}
  \includegraphics[height=.13\textwidth]{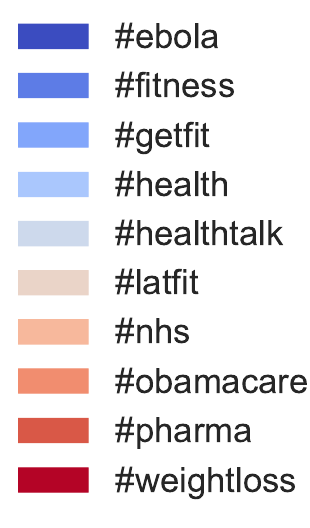}}
\caption[2-D Visualization of the tweets for the ten most popular hashtags for the three datasets.]{2-D Visualization of the tweets for the ten most popular hashtags for the three datasets. The tweets are represented by the 768-dimensional BERTag-covid embeddings. Note that the hashtags in the legend of each plot are sorted in alphabetical order rather than decreasing order of frequency as in Table~\ref{tab:ten-hashtags}. (figure best viewed in colour)}
\label{fig:clusters1}
\end{figure*}

As discussed above, compared to Word2Vec and FastText, when BERTag-base or BERTag-covid embeddings are used to represent hashtags, the average \hrRE s are consistently higher for all the datasets.  To illustrate how the clusters of tweets sharing the same hashtags look like, we pick the ten most popular hashtags in each dataset (see Table~\ref{tab:ten-hashtags}) and  project all the tweets for these hashtags to a 2D space using the t-SNE algorithm. Due to space limitation, we only show the t-SNE projection for the BERTag-covid embeddings.  Some details about Fig.~\ref{fig:clusters1} are listed below.

\begin{table}[!t]
\centering
\caption{The ten most popular hashtags and their frequencies}
\label{tab:ten-hashtags}
\begin{tabular}{|C{1cm}| L{12cm} |}
\hline \footnotesize
\textbf{Dataset} & 
\multicolumn{1}{c|}{\textbf{Popular hashtags and their frequencies}} \\ \hline\hline
\myCparbox{1cm}{AU Trends}     & \myparbox{12cm}{\#covid19 (81900),  \#coronavirusupdate (59514),  \#coronaviruspandemic (52163),  \#coronavirus (46290),  \#heabreakweather (27885),  \#covid2019 (20797), \#coronaoutbreak (17403),  \#corona (9801),  \#coronav (9150),  \#flattenthecurve (6595)} \\ \hline
\myCparbox{1cm}{US Trends}     & \myparbox{12cm}{\#heabreakweather (66262),  \#niallcarpool (16328),  \#coronavirusinkenya (10203),  \#1 (10089),  \#fridaythe13th (8442),   \#2 (5849),  \#traderjoes (4671), \#bernieisright (4578),  \#latelateniall~(4352),  \#covid19 (4345)}  \\ \hline
\myCparbox{1cm}{Health}        & \myparbox{8cm}{\#healthtalk (873),  \#nhs (768),  \#ebola (392), \#getfit (261),  \#latfit~(241),  \#obamacare (237),  \#weightloss (235),  \#health (228),  \#pharma (209), \#fitness (209)}  \\ \hline
\end{tabular} 
\end{table}

\begin{itemize}
    \item For the AU Trends dataset, as the tweets were crawled in the period when Covid-19 was widely discussed on Twitter, nine out of these ten hashtags are related to Covid-19,  except for  \#heabreakweather, which is related to the singer Niall Horan's studio album \textit{Heartbreak Weather} released on the same date. It is clear in Fig.~\ref{fig:clusters1}(a) that the tweets for this hashtag form a cluster on their own (the deep red dots) while the tweets for the other nine clusters all lump together as they are related to the same topic.     
    \item For the US Trends dataset (Fig.~\ref{fig:clusters1}(b)), the largest cluster is the tweets for the first hashtag \#heabreakweather. There is a great overlap between the two clusters for \#1 and \#2, which are common hashtags that Twitter users like to use when they want to claim they were the first person achieving certain endeavours. As expected, the clusters for the two Covid-related hashtags,  \#coronavirusinkenya and \#covid19, overlap significantly, while the tweets for \#bernieisright and \#fridaythe13th form two separate clusters.  Many tweets for the hashtag \#traderjoes (about Trader Joe's grocery store) had cross-topic discussions and they form scattered points, overlapping with other clusters.
    \item Different from the previous two datasets, the Health Dataset is a domain-specific dataset crawled in 2015. For this dataset (Fig.~\ref{fig:clusters1}(c)), the tweet clusters for most hashtags are well separated, e.g., \#ebola, \#obamacare, \#nhs, and \#pharma. However, due to the similarities of the topics discussed in the tweets, the \#health cluster has some overlap with \#healthtalk and \#fitness; the tweet cluster for \#weightloss overlaps with those for \#latfit and \#fitness, and it is also close to the cluster for \#getfit.     
\end{itemize}

We can see from the t-SNE projection that the hashtags represented using BERTag-covid (similarly for BERTag-base) are more closely related to the topics discussed in the tweets. Consequently, these embeddings give more relevant synonyms for hashtag recommendation evaluation, as confirmed also from the average \hrRE s shown in Fig.~\ref{fig:performance_hit_ratio_BERTag}.  
Our \#REval framework therefore uses either BERTag-base or BERTag-covid as the default embeddings for the BERTag module.

\smallskip

\noindent \textbf{\small(iv)~Effectiveness and correctness of the \hrRE\ formula.} We discussed in Section~\ref{sec:sematic-eval} why the list of synonyms should be constructed for the hashtags in the recommendation set $R$ only, as formulated in Eq.~\eqref{eq:hrREval}.  It is clear that all the curves in Fig.~\ref{fig:performance_hit_ratio_BERTag} have a steeper increase of the average \hrRE\ when the number of synonyms is small and only a gentle increase when more synonyms are included in the \hrRE\ calculation. This is what we expect to see in a good hashtag recommendation model. Of course, if we keep growing $k$ indefinitely, we will continue to see a gradual increase of \hrRE.  So, when comparing the performance of two hashtag recommendation models, one should consider the model that achieves the same \hrRE\ but with a smaller $k$ value as a better model. Our current framework allows the $k$ value to be fed to the \textit{Synonym and Thesaurus Construction} module (Fig.~\ref{fig:REval}).  To impose the criterion that the synonyms must meet a certain level of quality, a threshold value on the cosine distance used in kNN (Section~\ref{sec:knn}) can be easily added to the module.

\section{Conclusion and Future Work}
\label{Conclusion}
We have presented \#REval, a novel evaluation framework that measures the performance of hashtag recommendation models by incorporating hashtag synonyms. The framework automatically constructs thesauri of hashtags with their synonyms based on their similarity in the hashtag (and tweet) embedding space. 
In the paper, we have introduced BERTag, an internal module of \#REval that learns hashtag representations from clusters of tweets encoded using a pre-trained transformer-based model.
Our experiments show that \#REval with BERTag-base and BERTag-covid hashtag embeddings give better and more relevant synonyms than the Word2Vec and FastText embeddings do.  We have shown that the exact-matching method is inadequate for matching hashtags and therefore not suitable for evaluating hashtag recommendation models. Our proposed \hrRE\ formula helps to overcome this short-coming.

While we use BERTag embeddings in our \#REval framework because of the superior performance of the parent BERTweet embeddings, other improved hashtag embeddings in the future can be used in the BERTag module.  Our \#REval framework can be trained for different datasets to generate different dictionaries.  Once the training is done, the framework can be put in action for evaluation, i.e., the second and third modules do not need any fine-tuning.  We think that, with a large number of social media users today, having an evaluation framework such as \#REval is important. Our framework can also be adopted for other applications such as 
product recommendation extracted from messages posted by users. Besides, the resulting hashtags representations encoded using BERTag can be used in many applications such as topic identification, event detection, and information diffusion. 

Currently, the \hrRE\ formula considers the synonym list for any given hashtag as a set. Our future work is to extend the formula by considering it as an ordered list. By doing so, higher weights can be given to those synonyms appearing near the beginning of list, as they are closer to the hashtag. Different ways of ranking hashtags in the synonym list can also be investigated and compared.


\bibliographystyle{unsrtnat}
\bibliography{references}

\end{document}